\documentclass[12pt]{article}
\usepackage{graphicx}
\usepackage{epstopdf}
\usepackage{amsmath}
\usepackage{amsfonts}
\usepackage{amssymb}

\usepackage{graphicx}

\def\ee{\end{eqnarray}}

\def\eff{\mathrm{eff}}

\newcommand{\be}{\begin{eqnarray}}
\newcommand{\en}{\end{eqnarray}}
\newcommand{\ena}{\end{array}\right)}
\newcommand\eea{\end{eqnarray}}
\newcommand\bea{\begin{eqnarray}}
\newcommand{\invMpc}{Mpc$^{-1}\;$}

\usepackage{epsfig}
\usepackage{amsfonts}
\usepackage{amssymb}
\usepackage{graphicx}
\usepackage{bm}

\begin{document}





\vspace{5mm}
\vspace{0.5cm}
\begin{center}

\def\thefootnote{\fnsymbol{footnote}}

{\Large \bf Non-Gaussianities from Perturbing Recombination}
\\[0.5cm]
{\large  Leonardo Senatore$^{a,b,c}$, Svetlin Tassev$^c$ and Matias Zaldarriaga$^{b,c}$}
\\[0.5cm]

\vspace{.3cm}
{\normalsize { \sl $^{a}$ School of Natural Sciences, Institute for Advanced Study, \\Olden Lane, Princeton, NJ 08540, USA}}\\

\vspace{.3cm}
{\normalsize { \sl $^{b}$ Jefferson Physical Laboratory, Harvard University, Cambridge, MA 02138, USA}}\\

\vspace{.3cm}
{\normalsize {\sl $^{\rm c}$  
Center for Astrophysics, Harvard University, Cambridge, MA 02138, USA}}



\end{center}

\vspace{.8cm}

\hrule \vspace{0.3cm}
{\small  \noindent \textbf{Abstract} \\[0.3cm]
\noindent We approximately compute the bispectrum induced on the CMB temperature by fluctuations
in the standard recombination epoch. Of all the second order sources that can induce
non-Gaussianity during recombination, we concentrate on those proportional to the
perturbation in the free electron density, which is about a factor of $5$ larger than
the other first order perturbations. This term induces some non-Gaussianity by
delaying the time of recombination and by changing the photon diffusion scale. We find 
that the signal is not scale invariant, peaked on squeezed triangles with the smaller
multipole around the scale of the first acoustic peak, and that its size corresponds to
an effective $f_{\rm NL}\simeq -3.5$, which could be marginally detected by Planck if
both temperature and polarization are measured.

\vspace{0.5cm}  \hrule
\def\thefootnote{\arabic{footnote}}
\setcounter{footnote}{0}

\def\be{\begin{eqnarray}}
\def\ee{\end{eqnarray}}



\section{Introduction}

Non-Gaussianities in the Cosmic Microwave Background (CMB) have become very important in the last few years. On the observational side, there has been a huge improvement in the sensitivity with experiments such as WMAP. Upcoming experiments such as Planck will be sensitive to even smaller non-Gaussian signals. On the theoretical side, standard slow-roll inflation predicts an extremely small amount of non-Gaussianities \cite{Maldacena:2002vr}, undetectable by next generation experiments. However, several large modifications to the standard inflationary picture have been proposed both in the case of single field inflation \cite{Arkani-Hamed:2003uz,Alishahiha:2004eh,Shandera:2006ax,Chen:2006nt}, where all the models have been unified in an effective field theory description in \cite{Cheung:2007st}, and in the case of multi-field inflation \cite{Lyth:2002my,Zaldarriaga:2003my}. Recently even a consistent bouncing cosmology has been proposed \cite{Lehners:2007ac,Buchbinder:2007ad,Creminelli:2007aq}, which, though significantly less compelling than inflation, does predict a possibly detectable level of non-Gaussianities \cite{Creminelli:2007aq}. A detection  of primordial non-Gaussianities would therefore imply a radical departure from the standard and familiar slow-roll inflation scenario. Non-Gaussianities are probes of the interactions of the inflaton, and therefore contain an unprecedented  amount of information on inflation. If detected, we would have to abandon the standard slow-roll inflation picture, but would be left with new information to understand the dynamics of the inflaton.

Of course, non-Gaussianities in the CMB are not generated only by the inflaton, there are many contributions coming from the post-inflationary evolution. These can be roughly subdivided into effects which are important around or before recombination, and foreground contamination or secondary anisotropies generated after recombination. The latter include scattering secondaries, such as the thermal Sunyaev-Zel'dovich (SZ) effect \cite{SZ}  and the kinetic SZ or Ostriker-Vishniac (OV) effect \cite{OV}; and gravitational secondaries, such as weak lensing and the Rees-Sciama (R-S) effect \cite{RS}.
In order to be able to fully exploit upcoming CMB experiments, it is necessary to understand quantitatively the contributions from all these post-inflationary mechanisms. 

The effect of the secondaries on the CMB bispectrum has been already extensively analyzed in the literature. The thermal SZ effect and its frequency dependence are investigated in \cite{0002238}. The R-S effect is studied in \cite{9811252v2}. Gravitational lensing and the ISW effect are considered in \cite{1999PhRvD..60d3504S} and \cite{9811251v2}. The OV effect is analyzed in e.g. \cite{astro-ph/0511060v1} and \cite{0610007v2}. The signal-to-noise for upcoming experiment generated by these secondaries is very large and it will have to be taken into account when looking for primordial non-Gaussianities. 

Let us now consider non-Gaussianities generated from around recombination and concentrate on the 3-point function or bispectrum. Due to translational invariance, the bispectrum in Fourier space is a function of three wavevectors which must add up to zero forming a closed triangle. For the squeezed triangle limit, where one of the modes is well outside the horizon, the bispectrum generated around recombination has been obtained exactly  in \cite{2004PhRvD..70h3532C}. Oversimplifying, we can  imagine that the non-Gaussianities are given by the following relation between the observed curvature perturbation $\zeta$ and a gaussian random variable $\zeta_g$:
\be
\zeta(\vec x)=\zeta_g(\vec x)-\frac{3}{5}f_{\rm NL}^{\rm loc.}\left(\zeta_g(\vec{x})^2-\langle\zeta_g(\vec{x})^2\rangle\right)\ ,
\ee 
where $f_{\rm NL}^{\rm loc.}$ is a measure of the level of non-Gaussianity, and the superscript loc., which stands for local, refers to the above local-in-space structure of the non-Gaussiantiy. Then, in \cite{2004PhRvD..70h3532C} it was found that the non-Gaussianity in these squeezed triangles  corresponds to an $f_{\rm NL}^{\rm loc.}<1$, which is undetectable even by Planck.

The purpose of this paper is to go beyond this simplified limit, and take a step further in the computation of the non-Gaussianities generated around recombination for all triangles. The full calculation of the bispecturm is a very hard task. 
In order to obtain it, one is forced to solve the second order Boltzmann and Einstein equations, which are very complicated. We give those in a companion paper \cite{inprep1}, following and  correcting the equations given in \cite{bartolo} and \cite{bartolo2}, and including the effects of the free electron number density perturbation. This we calculated at first order in the approximation of the Peebles' effective 3-level atom \cite{peebles}. In \cite{inprep1} we showed that around recombination, the amplitude of the perturbations to the free electron density $\delta_e$ is enhanced by a factor of $\sim5$ relative to the baryon density perturbations due to the relatively small timescale of recombination~\footnote{As usual, we denote the change in a given quantity $q$ by $\delta q$, while the fractional change we denote by $\delta_q\equiv \delta q/q$}. 

Without accounting for the enhancement of $\delta_e$, one very naively expects that the second order evolution will lead to a CMB bispectrum with an $f_{\mathrm{NL}}^{\rm loc.}\sim 1$, which for an experiment like Planck corresponds to a signal-to-noise ratio of $\sim0.3$ (see for example \cite{2004PhRvD..70h3005B}). Consequently, we may expect that the enhanced $\delta_e$ will lead to an observable three-point signature in the CMB corresponding to an $f_{\rm NL}^{\rm loc.}\sim 5$. In this paper, we therefore  concentrate only on those non-linear perturbations which are induced by $\delta_e$. 

Notice that, as long as the tight-coupling between photons and baryons holds, the CMB temperature does not depend on the electron number density $n_e$, and the contribution from $\delta_e$ to the second order CMB anisotropies vanishes. This means that the effect of $\delta_e$ is most important during recombination. 

At first order, the solution for the CMB temperature anisotropies $\Theta\equiv \delta T/T$, once expanded in multipoles, is given as an integral over the first order source $S^{(1)}$ and the photon visibility function $g(\eta)$ \cite{cmbfast}. For a mode of wave vector $\vec k$, the solution reads:
\be\label{firstorder}
\Theta_l^{(1)}(k,\eta_0)\sim\int_0^{\eta_0} d\eta g(\eta) S^{(1)}(k,\eta)j_l[k(\eta_0-\eta)]\ ,
\ee
where $j_l$ is the spherical Bessel function, $\eta$ is conformal time and $\eta_0$ corresponds to $\eta$ of today. The order in perturbation theory is given as a superscript in parenthesis and will be dropped whenever that does not cause confusion. If $S^{(1)}$ is expressed using the photon monopole, dipole, and quadrupole without accounting for photon diffusion, then including the effects of photon diffusion approximately amounts to replacing $S^{(1)}\to S^{(1)} \exp[-k^2/k_D^2]$, where $k_D$ is the photon diffusion scale (see for example Section 8.5 of \cite{Dodelson}). 

The electron density $n_e$ enters only in the visibility function and the diffusion scale, which means that $\delta_e$ multiplies the first order source. Thus, any perturbations to $n_e$ will affect $\Theta$ at second order only. This can be understood also intuitively.  In a homogeneous universe, before, during and after recombination the radiation temperature decreases as $a^{-1}$, $a$ being the scale factor, irrespective of the electron density. This means that a perturbation to the electron density changes the position of the last scaterring surface and the mean free path of the photons before recombination, but not the observed radiation temperature. At first order we perturb $n_e$ and keep the other quantities unperturbed, and therefore the CMB anisotropies are not affected by $\delta_e$ at this order. 

From (\ref{firstorder}), we expect that the expression for the second order temperature anisotropies after including the perturbation to $n_e$ will be schematically given by~\footnote{In the paper we expand the perturbations as $$f(\vec x,p^i,\eta)=f^{(0)}(p^i,\eta)+f^{(1)}(\vec x,p^i,\eta)+\frac{1}{2}f^{(2)}(\vec x,p^i,\eta)\ .$$}
\be\label{guess}
\Theta^{(2)}_{lm}(k,\eta_0)\sim 2 \int_0^{\eta_0} d\eta g(\eta) \left[{\delta_g}+2 \frac{k^2}{k_D^2}\delta_{k_D}\right] S^{(1)}(k,\eta)j_l[k(\eta_0-\eta)]\ ,
\ee
where $\delta_g$ and $\delta_{k_D}$ are the fractional perturbations to the visibility function and diffusion scale. When later in the paper we perform the full calculation, we will see that the above schematic equation is not far from the right expression.

The paper is organized as follows. In sec.~{\ref{sec:estimates}} we elaborate on eq.~(\ref{guess}) and provide simple estimates of the induced bispectrum. Then, in sections~\ref{sec:second_order} and~\ref{sec: 3-point function}, we go on and present the full calculation of the bispectrum concentrating on modes inside the horizon so that we expect to be able to neglect second order metric perturbations. We summarize the result of the analytical calculation in sec.~\ref{sec:summary of analytical}. In sec.~\ref{sec:resultsand discussion}, we show the results of the numerical integration and describe the induced signal. We conclude in sec.~\ref{sec:summary}. In App.~\ref{appendix:A} we summarize some useful formulas about rotation invariance.

\section{(Not so) simple estimates\label{sec:estimates}}

\subsection{Mechanisms generating $\Theta_l^{(2)}$ \label{sec:mechanism}}

Let us begin to analyze the mechanisms  by which perturbations in the number of free electrons can induce a bispectrum  in the CMB.  We restrict our analysis to modes  which are faster than the horizon scale at recombination ($k\gtrsim0.005 $ \invMpc), and slower than the scale associated to the width of the visibility  function ($k\lesssim0.27 $ \invMpc). This is the region from where we expect that a large three point function can be generated. For modes which are out of the horizon at recombination, the calculation of \cite{2004PhRvD..70h3532C} which is valid in the limit in which one of the modes is outside of the horizon and the other two modes are much faster than the first one, shows that the expected signal is equivalent to $f_{\rm NL}^{\rm loc.}< 1$, which is very small. Similarly, we do not expect a large signal coming from configurations where all of the modes are out of the horizon and of comparable size. We do not expect a high signal also from modes with $k\gtrsim 0.27 $ \invMpc because in this case the mode is faster than the timescale of recombination, and there is effectively an averaging out of the perturbation \cite{inprep1}. Furthermore, these very high $k$ modes lie outside the reach of experiments such as Planck.

As eq.~(\ref{guess}) suggests, $\delta_e$ will affect the CMB bispectrum in three different ways: by perturbing the physical time at which recombination takes place, by perturbing the photon diffusion scale, and by perturbing the probability  for the CMB photons to originate 
from different times within the recombination era. This last mechanism could be relevant because diffusion damping during recombination smoothes out the perturbation associated with a given mode on a time scale corresponding to approximately the period of oscillation of the mode itself.  The later in time photons originate from (even if an order one Hubble time after the peak of the visibility function), the more the anisotropy is suppressed. We will find that this effect is subdominant.

The different contributions can be extracted by writing eq.s~(\ref{firstorder}) and (\ref{guess}) as
\be\label{guessold}
\Theta^{(1+2)}_{lm}(k,\eta_0)&\sim&\int_0^{\eta_0} d\eta g^{(0)}(\eta)(1+\tilde{\delta_g}) \exp[-k_s^2/k_D^2(\eta+\delta\eta_r(\vec k_e),n_e+\delta n_e(k_e))]\nonumber\\ 
&&\times S^{(1)'}(\eta+\delta\eta_r(\vec k_e)) j_l[k_s(\eta_0-\eta-\delta\eta_r(\vec k_e))]\ ,
\ee
where $S^{(1)'}$ is the first order source without the exponential damping; $n_e$ is the free electron number density; $c_s$ is the sound speed; $k_s$ is the wavenumber of the source; $k_e$ is the wavenumber of $\delta_e$; $g^{(0)}$ is the unperturbed visibility function; and a convolution such that $\vec k_s+\vec k_e=\vec k$ is implicit.  $\delta\eta_r(\vec k_e)$ is the perturbation to the position of the peak $\eta_r$ of the full visibility function
\be
g^{(0)}(\eta)+\delta g(\eta, k_e)\ ,
\ee
where $\delta g$ is the perturbation to the visibility function. Still in eq.~(\ref{guessold}), we have split the perturbation to $g$ into two pieces: 
\be\label{deltaG}
\delta g=-\dot g(\eta) \delta\eta_r(\vec k_e)+\tilde{\delta g}(\vec k_s,\vec k_e)\ .
\ee
The first piece gives approximately the time shift of the (otherwise unperturbed) visibility function, while $\tilde{\delta g}$ takes into account the modification of its shape. 

By shifting the variable of integration, in eq.~(\ref{guessold}) we have moved $\delta\eta_r$ inside the first order source. In the limit in which the visibility function can be roughly thought as a $\delta$-function centered at $\eta_r$, the effect of $\delta\eta_r$ is to evaluate $j_l$ and all the time dependent quantities in the first order source term at a shifted time.
After this time shift  has been taken into account, another possibly important effect of the perturbation of $g$ is its change in shape. As we will later see, this effect will turn out to be unimportant, which we will show in the estimates that follow. Even in the case in which a $\delta_e$ mode does not change the position of the peak of the visibility function at recombination, it can still distort its shape, by for example making the width smaller. Because in the time scale of the order of a few widths of the visibility function the source decays relevantly (even for relatively low $k_s$), the actual location around recombination from where most  of the CMB photons originate, is relevant in determining  the magnitude of the CMB anisotropy. If there is more probability in the central part of recombination, then the anisotropy is larger, while in the opposite case, it is smaller (see Fig.~\ref{fig:deltaarea}). Clearly, this effect vanishes in the limit where the diffusion damping scale goes to infinity, because in this case the
size of the source is independent of the place from where the photon originates. It also vanishes in the limit where the  $\delta_e$ mode is oscillating on a time scale either much faster than the diffusion damping scale of the source mode, because in this case an averaging occurs, or much slower, because in this case it will just shift the peak of the visibility function, but not change its shape. Of course, for a generic $\delta_e$ perturbation, a part of the above effect is taken into account by the shift in the position of the peak $\eta_r$ of the visibility function. In order to take into account the effect due to the change in its shape, we define a perturbation to the  `effective Area' of the visibility function after the time shift has been subtracted:
\be\label{eq:effectivearea}
\delta \mathrm{Area}=\frac{\int d\eta g^{(0)}(\eta)\tilde{\delta_g}(\eta)\exp[-k_s^2/k_D^2(\eta)]}{\int d\eta g^{(0)}(\eta)\exp[-k_s^2/k_D^2(\eta)]}\ ,
\ee
where, as explained,  the weighting by the diffusion damping takes approximately into account the importance of the location of the probability to create a CMB photon, for the resulting size of the CMB anisotropy.

Before we procede we would like to have an expression for the perturbation to the visibility function, $\delta_g$, wich includes the oscillatory character of $\delta_e$. This can be taken into account if one perturbs $\delta_g$ not simply with the amplitude, $\delta_e$, but with a multipole expansion of $\delta_e(k_e,\eta)\exp(i\vec{k}_e\cdot \vec{x} )$. The angle-independent part of the multipole expansion of $\exp(i\vec{k}_e\cdot\vec{n}(\eta_0-\eta))$ is $\propto j_{l}(k_e(\eta_0-\eta'))$. We therefore use the following effective perturbation to $g$:
\be\label{deltageff}
j_{l_{\eff}}(l_\eff)\delta g (k_e,\eta)=g^{(0)}(\eta)\left(
\delta_e(k_e,\eta)j_{l_\eff}(k_e(\eta_0-\eta))-\int^\eta_{\eta_0}d\eta' \dot\tau(\eta')\delta_e(k_e,\eta')j_{l_\eff}(k_e(\eta_0-\eta'))
\right)\ ,
\ee
with $l_\eff\equiv k_e(\eta_0-\eta_r)\nonumber$, where we have pertured $g$ with $\delta_e(k_e,\eta) j_{l_\eff}(k_e(\eta_0-\eta))$.  This choice of perturbation approximates the result of our full treatment (cf. eq.~(\ref{eq:B_total_simple})). 

The total perturbation to $g$ at late times is large because there is still quite a bit of optical depth after recombination ($\tau=0.04$ between $\eta=360$\invMpc and $\eta_0$) and $\delta_e$ grows once it starts falling in the CDM potential wells \cite{inprep1}. This could naively give a large effect. However, in order to accumulate a good fraction of this optical depth the photons need to travel a large distance comparable to $\eta_0$. This is possible only for $\hat k$ being to a very good approximation perpendicular $\hat n$, since otherwise the oscillating $\delta_e$ averages out along the line of sight. This implies that for a fixed $|\vec k|$ and $l$, there are very few $\hat k$ directions which contribute to the signal. This effect is taken into account in eq.~(\ref{deltageff}) by the spherical Bessel function. This suppression of $\delta g$ at late times renders the $\delta$Area effect small as we will see below.

\begin{figure}[h!]
\begin{center}
\includegraphics[width=8cm]{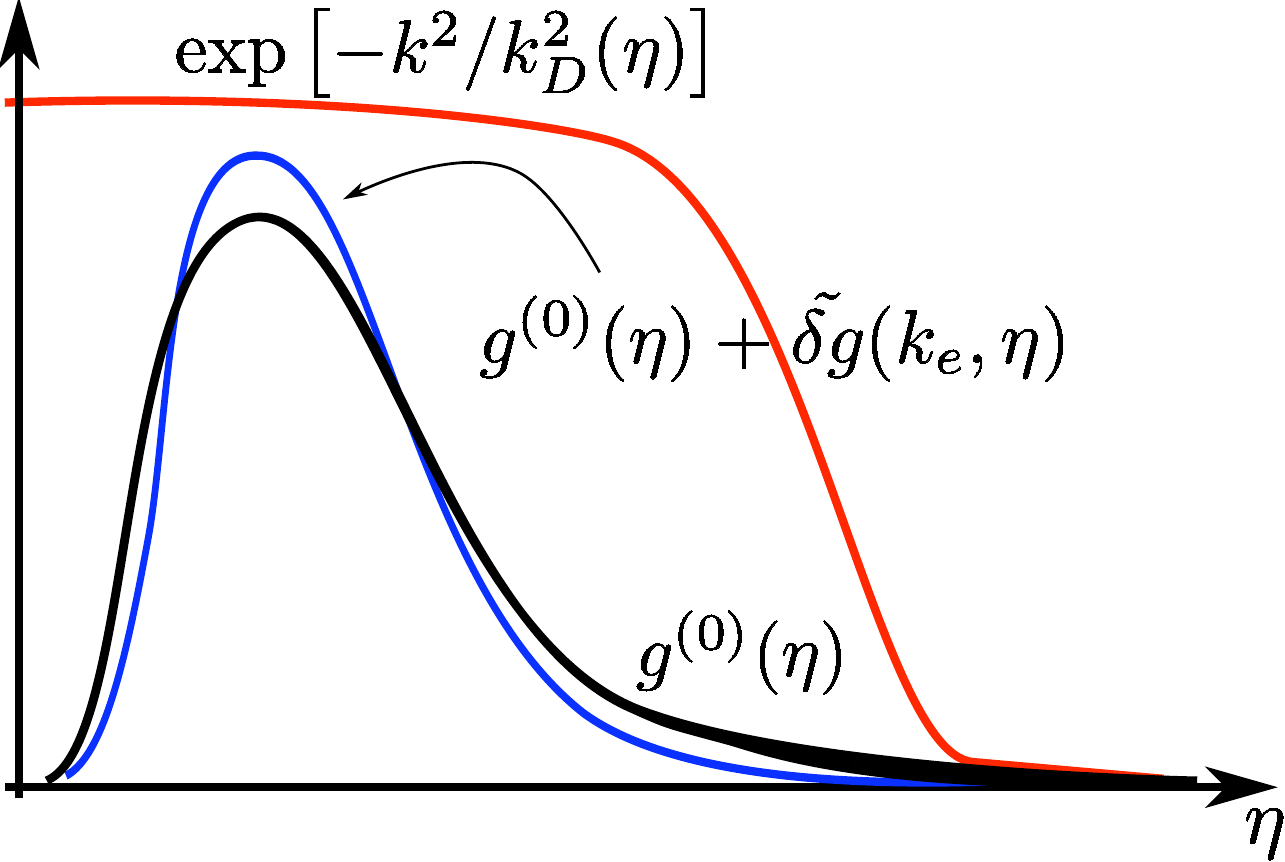}
\end{center}
\caption{\small Schematic representation  of the effect of a perturbation $\tilde{\delta g}$ to the shape of the visibility function for the case high-$k_e$ and low-$k_s$. Notice that the location of the peak is unchanged, so that $\delta\eta_r=0$. However, $g$ is much more peaked around the central value than in the homogenous case. This means that more CMB photons originate from the region near the peak, where the power of the mode is not yet damped (in red we plot the diffusion damping). This means that the resulting CMB anisotropy is larger. The shaded black region of $g^{(0)}$ represent how much the  `Effective Area' in the homogeneous case differs from one and it is a measure of how many CMB photons originate from the region where the mode is already beginning to be damped. We see that $g^{(0)}+\tilde{\delta g}$ has a larger `Effective Area', implying a positive $\delta$Area and the possibility for a larger CMB anisotropy. 
}\label{fig:deltaarea}
\end{figure}

From eq.~(\ref{guessold}), we obtain an approximate expression analogous to eq.~(\ref{guess})  that  we can investigate semianalytically and that explicitly differentiates between the different mechanisms generating $\Theta_l^{(2)}$:
\be\label{guesstwo}
\Theta^{(2)}_{lm}(k,\eta_0)&\sim& 2 \int_0^{\eta_0} d\eta g(\eta)\exp[-k_s^2/k_D^2(\eta)] \left[2 \frac{k_s^2}{k_D^2}\delta_{k_D}+2 \frac{k_s^2}{k_D^2}\delta_{k_g} +{\delta \mathrm{Area}}+ \delta\eta_r\; \partial_\eta\right] \nonumber\\
&&\times S^{(1)'}(k_s,\eta)j_l[k_s(\eta_0-\eta)]\ ,
\ee
where $\delta_{k_g}\equiv \delta\eta_r \dot{k}_D/k_D$ is the perturbation to $k_D$ due to the timeshift $\delta\eta_r$ in the visibility function which makes $k_D$ to be evaluated at a slightly different time than in the unperturbed universe $\big($notice that $k_D(\eta)\propto\sqrt{n_e(\eta)}$ $\big)$, while $\delta_{k_D}$ measures the perturbation to $k_D$ due to $\delta n_e$ directly~\footnote{Notice that we denote by $\delta k_D$ the full perturbation to $k_D$, due to both of the effects, and not just to $\delta n_e$.}. It is quite straightforward to realize that, in the limit in which the timescale over which the source varies is much longer than the width of the visibility function, the above equation gives an approximate expression for the dependence of $\Theta^{(1+2)}$ on the $\delta$Area. This expression is expected to be correct up to order one corrections, which is enough for the purposes of the present section. 
As we will later show (see later in eq.~(\ref{blllfirst})), the 3-point function will be sensitive to the variation of the angular power spectrum $C_l$ with respect to the quantities perturbed by $\delta_e$. In the case of the phase shift $\delta\eta_r$, this gives rise to a term proportional to $(-c_s k_s \delta{\eta_r}) \sin(c_s k_s\eta_r)$, which, once integrated in Fourier space to obtain the $C_l$'s, gives rise to approximately an effect of order $(-c_s k_s \delta{\eta_r})/3$. In the regime of interest, the factor of $\sim1/3$ takes approximately into account of the difference from integrating over $k_s$  the source expression that now contains  $\cos(c_s k_s\eta_r)\sin(c_s k_s\eta_r)$ instead of $\cos^2(c_s k_s\eta_r)$, in both cases multiplying other $k_s$ dependent terms. We verified that the perturbation to $\eta$ entering in $j_l$ leads to a negligible contribution, and we therefore drop it here.

\begin{figure}[h!]
\begin{center}
\includegraphics[width=15cm]{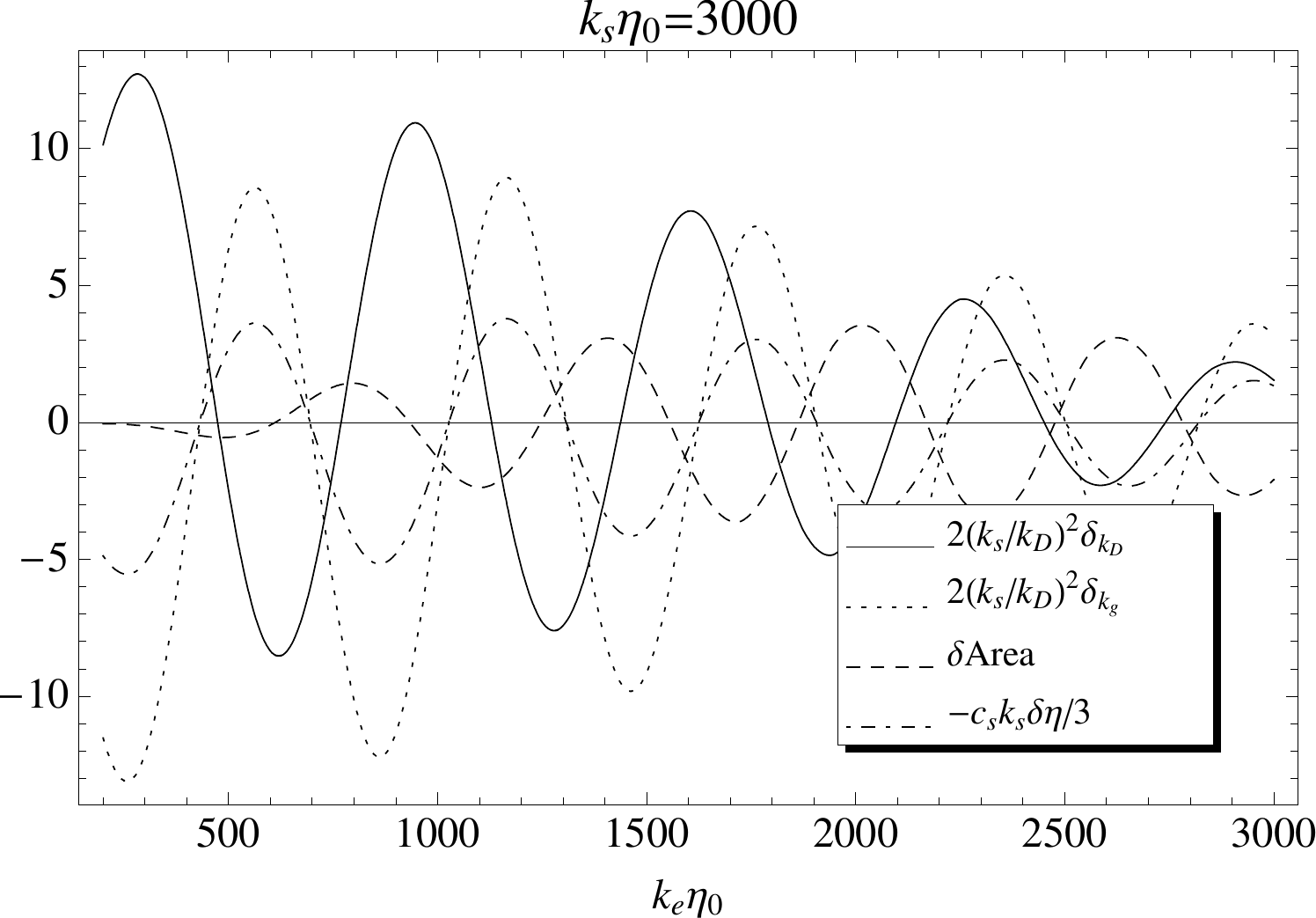}
\end{center}
\caption{\small \small  The first-order perturbations responsible for the bispectrum generated from recombination. The computation was done in synchronous gauge, and the normalization of the perturbations is the one in CMBFAST \cite{cmbfast}-- on superhorizon scales $\zeta=1$.  $\delta_{k_D}$ represents the perturbation to the diffusion scale at the peak of the visibility function ($\eta\simeq 288\,$Mpc) due to $\delta n_e$; $\delta_{k_g}$ is the perturbation to $k_D$ due to $\delta\eta_r$ -- the shift in the position of the last scattering surface. $\delta\mathrm{Area}$ approximately represents the perturbation to the probability that a photon originates from the last scattering surface before the perturbation  decays due to diffusion damping. The timeshift $\delta\eta_r$ also gives rise to a change in the phase at which we evaluate the first order source, which is schematically given by $(-c_s k_s \delta \eta_r/3)$. The plot is for $k_s\eta_0=3000$. From the plot we can see that the largest contribution to the second order temperature anisotropies for the given $k_s$ is from $k_e\eta_0\sim 200$, i.e. from around the first acoustic peak. Note that the two largest contributions to $\Theta_l^{(2)}$ -- the perturbations of $k_D$ coming from $\delta_e$ and the change in the position of the last scattering surface, partially cancel each other. 
}\label{fig:kDL}
\end{figure}

\begin{figure}[h!]
\begin{center}
\includegraphics[width=15cm]{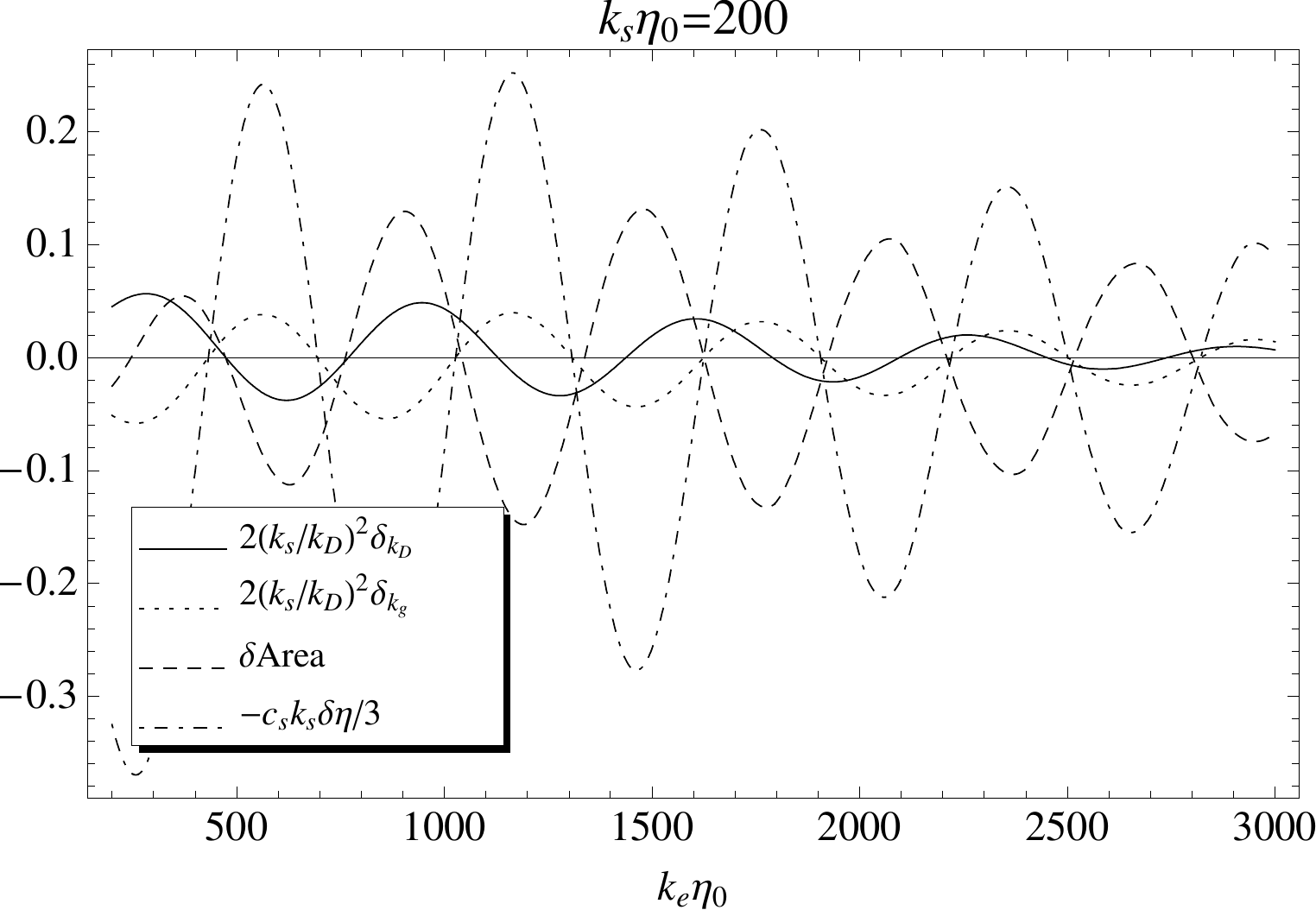}
\end{center}

\caption{\small \small  The same as in Fig.~\ref{fig:kDL} but for $k_s\eta_0=200$. As one can see, all effects for high $k_e$ suffer from some kind of suppression as described in the text.
}\label{fig:kDH}
\end{figure}

Using our results from \cite{inprep1} for $\delta_e$, in Figures \ref{fig:kDL} and \ref{fig:kDH} we plot the above four first-order quantities. All calculations were done in the synchronous gauge. 

When we compute the bispectrum, we compute the correlation function of three $\Theta$ modes. In order for this to be non zero, one of the modes has to be taken at second order, obtaining an expression  of the form  (neglecting multipole indices for simplicity)
\be\label{eq:simple_estimate}
\langle\Theta^{(2)}(\vec k_1)\Theta^{(1)}(\vec  k_2)\Theta^{(1)}(\vec k_3)\rangle\sim\langle\delta_e^{(1)}(\vec  k_e)\Theta^{(1)}(\vec k_s)\Theta^{(1)}(\vec k_2)\Theta^{(1)}(\vec k_3)\rangle \ ,
\ee
where $\vec{k}_e+\vec{k}_s=\vec k_1$. As shown in Fig.~\ref{fig:triangle}, when computing the expectation value, we pair each first order perturbation with the others. This forces $\vec k_2=-\vec k_e$ and $\vec k_3=-\vec k_s$, while the sum of the three momenta $\vec k_1$, $\vec k_2$ and $\vec k_3$ must form a closed triangle by translation invariance. As we can see from Fig.~\ref{fig:kDL}, for a given  high $k_s$, the signal grows as we make $k_e$ lower~\footnote{This is true only until $k_e$ is within the horizon. For $k_e$ outside of the horizon $\delta_e$ becomes irrelevant. In this paper we refer to low-$k$ modes as modes  that project on the sky as multipoles with $l\sim 200$, while by high-$k$ modes we mean modes that give rise to multipoles of order thousands. Similarly by a low-$l$ we mean an $l\sim 200$ and by high-$l$ an $l\sim$ thousands.}. In the opposite case, for a low $k_s$, we see in Fig.~\ref{fig:kDH} that the signal never grows as we make $k_e$ larger. This makes us suspect that most of the signal comes from combining a low $k_e$ with a first order source term at high $k_s$. This is actually true, as we will confirm in the full calculation: the signal is peaked on squeezed triangles as the one shown in Fig.~\ref{fig:triangle}.

\begin{figure}[h]
\begin{center}
\includegraphics[width=9cm]{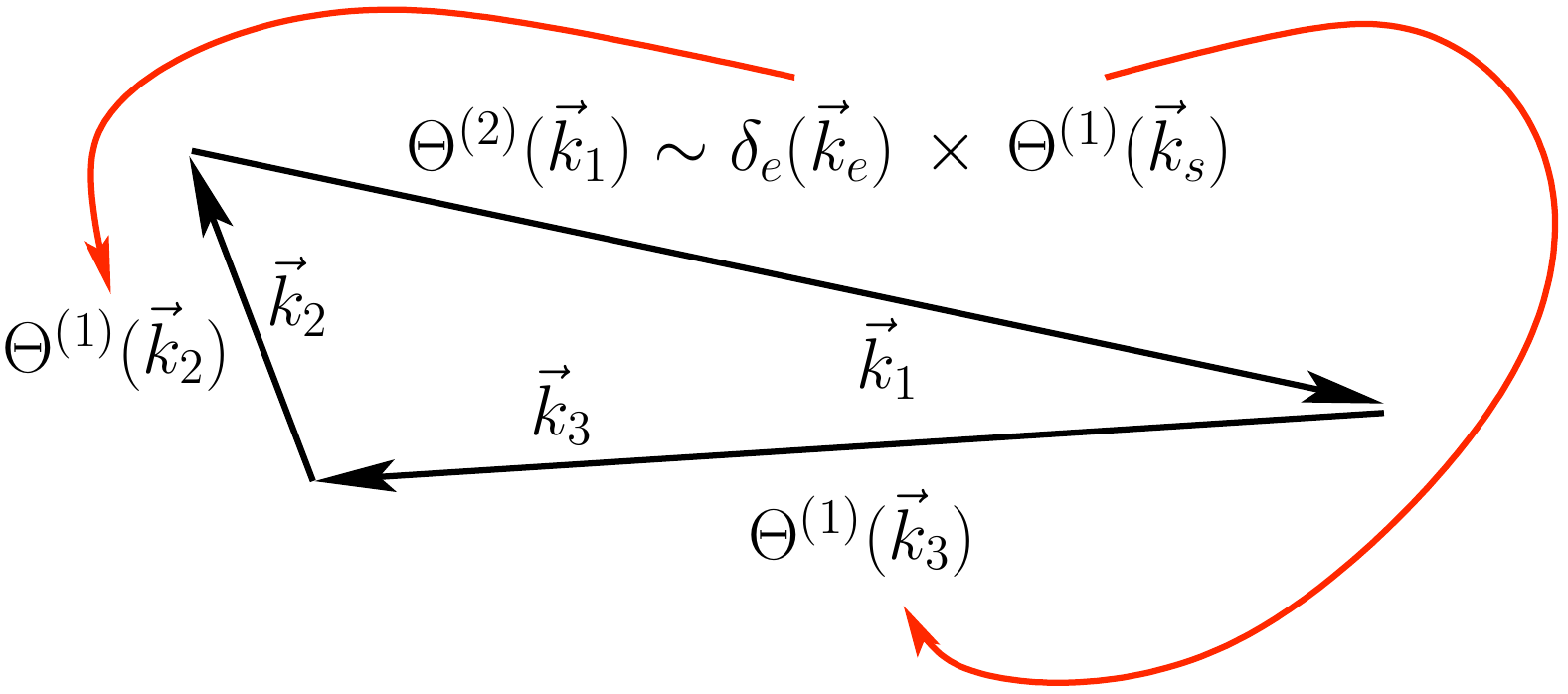}
\end{center}
\caption{\small \small  When computing the bispectrum, we take the expectation value of three $\Theta$ modes, and one of them has to be taken at second order in order not to have a null result. Each of the two first order perturbations contained in the second order mode, approximately $\delta_e$ and $\Theta^{(1)}$, need to be matched with one of the two first order $\Theta$ modes, forcing $\vec{k}_e=-\vec k_2$ and $\vec{k}_s=-\vec k_3$, where $\vec k_1=\vec k_e+\vec k_s$. The sum of  $\vec k_1$, $\vec k_2$ and $\vec k_3$  must be equal to zero, so that the three wave vectors form a closed triangle. The same is done for all the symmetric combinations. 
}\label{fig:triangle}
\end{figure}

The two largest contributions to the combination of a low $k_e$ and a high $k_s$ mode  (see Fig. \ref{fig:kDL}) are given by the perturbations to the diffusion scale, $\delta_{k_g}$ and $\delta_{k_D}$, which are comparable in magnitude but out of phase. These quantities are not enhanced by themselves, but they are multiplied by a factor of $2k_s^2/k_D^2$ in the expression for the second order temperature anisotropies (\ref{guesstwo}), which boosts their individual effects on the bispectrum, although their net effect is nearly completely cancelled. This cancellation can be easily understood since $n_e$ is larger in overdensities. This means that in the presence of a low-$k_e$ overdensity, the high $k$ modes (implying high $k_s$) will be less supressed due to the reduced diffusion scale (i.e. positive $\delta_{k_D}$), but they will have more time to decay (i.e. negative $\delta_{k_g}$) until recombination takes place. This can be understood also in the following way. In the limit in which the $\delta_e$ mode is much slower than the timescale of recombination ($k\lesssim 0.1$ Mpc$^{-1}$),  we expect that recombination happens in the same way as in the unperturbed universe,  just a bit time translated. For this reason, around recombination, $n_e$ will be just the time translation of its unperturbed value.  Since $k_D^2\sim \sigma_T n_e(\eta)a(\eta)/\eta$, where $\sigma_T$ is the Thomson cross section, after having taken into account of the time translation of $n_e$, the piece proportional to $\delta n_e$ disappears, and one is left only with a perturbation to the scale factor and to $\eta$, which give a very small effect. Notice that in this same regime, $n_e$ is determined by the local value of $n_b$ (or equivalently of the temperature) with the same relationship as in an unperturbed universe: $\delta n_e\simeq \delta n_b\, \dot n_e/\dot n_b$. As one can see in Fig.~5 of \cite{inprep1}, this is in fact the case  even for modes very well inside the horizon, implying that the cancellation between $\delta_{k_D}$ and $\delta_{k_g}$ begins to be milder only for relatively high $l$ modes. 

This is not however the whole effect in the regime of  low $k_e$ and high $k_s$. Still from Fig.~\ref{fig:kDL}, we expect that for this combination of $k$'s, the phase shift due to $\delta\eta_r$ in the source should also play an important role for the bispectrum. As we will later show, the full calculation involves an integral in time, which suppresses the effect from the phase shift with respect to the naive expectation from Fig.~\ref{fig:kDL}, due to the fact that in this case the integrand changes sign. Finally, in this regime, the perturbation to the area is not significant, which tells us that, as expected, for low $k_e$, $\delta_g$ is well approximated by just a timeshift in $g$. 

The contributions to $\Theta_l^{(2)}$ in eq.~(\ref{guesstwo}) from the combination of a low $k_s$ source mode and a high $k_e$ $\delta_e$ are shown in Fig.~\ref{fig:kDH}. Since $k_s$ is small, the factor $2k_s^2/k_D^2$ supresses the contribution of $\delta_{k_D}$ and $\delta_{k_g}$ in this case. Also the time shift is quite small. The effect that comes from the perturbation of the shape of the visibility function and therefore to the probability for the CMB photons to originate  from different times within the recombination era is also shown in Fig.~\ref{fig:kDH}. This is what we called $\delta {\rm Area}$ and described at the beginning of the section. This effect does not become important even for very high $k_e$ that change appreciably within the width of the visibility function, due to cancellation effects as described around eq.~(\ref{deltageff}). 

For a high $k_e$ perturbation of order 1, we can estimate the effect it will have on the bispectrum. Starting from eq.~(\ref{eq:simple_estimate}), we can write the ratio of the induced bispectra in the two different $(k_e,k_s)$ limits as
\be\label{highkorderone}
\frac{({\rm perturbation}(k_{\rm high})\sim 1) \Theta(k_{\rm low}) \times \Theta(k_{\rm high})\times \Theta(k_{\rm low}) }
{2 \frac{k_{\rm high}^2}{k_{ D}^2}\delta_{k_D}(k_{\rm low})\Theta(k_{\rm high}) \times \Theta(k_{\rm high})\times \Theta(k_{\rm low}) }\sim \frac{1}{13\, {\rm Exp}\left[{-k_{\rm high}^2/k_{ D}^2}\right]}\sim 0.5 \ ,
\ee
where $k_{\rm high}\eta_0\simeq 3000$ and $k_{\rm low}\eta_0\simeq 200$, and where we took the value of $\delta_{k_D}$ from Figure~\ref{fig:kDL}. 
Notice that the reason why the above ratio is order one is not because the high-$k_e$ perturbation is large, but because the ratio involves $\Theta(k_{\rm low})/\Theta(k_{\rm high})$ which is enhanced by the exponential damping. This means that not-enhanced first order perturbations that are not exponentially suppressed at high $k$, might give rise to an enhanced bispectrum. All of our high-$k_e$ perturbations are suppressed, as is the case for $\delta$Area due to cancellation effects, and therefore we do not expect a large signal to noise from this regime.

\subsection{Estimates for the bispectrum \label{sec:estimate_bispectrum}}
As discussed in Section \ref{sec:mechanism}, we expect a possible enhancement to the bispectrum from recombination only for modes with $l$ between approximately 100 and 3000. A simple approximate formula for the 3-point function can be obtained by noticing that expression (\ref{guesstwo}) for the second order $\Theta_{lm}^{(2)}$ is simply given by the product of a perturbation to a quantity `$X$' induced by $\delta_e$ times the derivative of the first order source with respect to $X$. Therefore, in this case, in the limit in which $\delta_X$ does not vary appreciably during the width of the visibility function, we can approximately write the 3-point function of the multipoles of the CMB anisotropy $a_{lm}$'s as
\be
\langle a_{l_1m_1} a_{l_2m_2}a_{l_3m_3}\rangle={\cal G}^{m_1m_2m_3}_{l_1l_2l_3}i^{l_1+l_2+l_3}b_{l_1l_2l_3}\ ,\ 
\label{blllfirst}
\ee
where
\be\label{eq:approx_bispectrum}
b_{l_1l_2l_3}=\langle a_{l_1m_1}\delta_{X,l_1m_1}^* \rangle\frac{1}{2}\frac{\partial C_{l_2}}{\partial \ln X}+ 5\ {\rm perm.}\ ,
\ee
 ${\cal G}^{m_1m_2m_3}_{l_1l_2l_3}$ is Gaunt integral, and the factor $i^{l_1+l_2+l_3}$ comes from our choice of phase for the $a_{lm}$'s (see eq.~(\ref{eq:alm})  below). In the flat sky limit the Gaunt integral just reproduces $(2\pi)^2\delta^2(\vec l_1+\vec l_2+\vec l_3)$. We will verify explicitly this formula when later we perform the full calculation. In the above equation we defined
\be
\delta_{X,lm}\equiv\frac{(-i)^{-l}}{2\pi^2}\int d^3kY^*_{lm}(\hat k)\zeta(\vec k) \delta_X(k,\eta_r)j_l(k\,\eta_0)\ ,
\ee
in analogy with 
\be\label{eq:alm}
a_{lm}\equiv\frac{(-i)^{-l}}{2\pi^2}\int d^3kY^*_{lm}(\hat k)\zeta(\vec k) S^{(1)}(k,\eta_r)j_l(k\,\eta_0)\ ,
\ee
where $\eta_r$ denotes $\eta$ at recombination. Here we have introduce the primordial curvature perturbation $\zeta(\vec k)$. In this case the first order quantities like $\delta(k,\eta_r)$ or $S^{(1)}(k,\eta_r)$ are meant as transfer functions. 

We can perform the angular integration in the expectation value in (\ref{eq:approx_bispectrum}) to obtain
\bea\label{blllsecond}
b_{l_1l_2l_3}&=&\left[\frac{2}{\pi}\left(\int dk k^2 P(k)\Theta_{l_1}(k,\eta_0)j_{l_1}(k\,\eta_0)\delta_X(k,\eta_r)\right)\times\frac{1}{2}\frac{\partial C_{l_2}}{\partial \ln X}+\right. \\ \nonumber
&&\left.\frac{2}{\pi}\left(\int dk k^2 P(k)\Theta_{l_2}(k,\eta_0)j_{l_2}(k\,\eta_0)\delta_X(k,\eta_r)\right)\times\frac{1}{2}\frac{\partial C_{l_1}}{\partial \ln X}\right]+2\ \mathrm{perm.}\ .
\eea
When later we  perform the full calculation, we will see that the above equation is a good approximation in the limit in which $\delta_X$ does not vary within the width of the visibility function, which corresponds to multipoles of order a few thousands. 

The above expression can be evaluated numerically with the approximate expression for $j_l$ given in eq. 9.71 of \cite{2005pfc..book.....M}, and using the approximation for the monopole and dipole of the first order source, given in eq.~8.24 and 8.26 of \cite{Dodelson} (where we drop the integral contributions for simplicity). We also include the exponential damping. The sign of the reduced bispectrum is chosen such that a positive bispectrum in the local model implies a negative $f_{\rm NL}^{\rm loc.}$ in the Sachs-Wolfe regime, and vice versa.  Of course this expression will not be very accurate,  due to the approximate treatment of diffusion damping in the first order source, to the additional averaging across the visibility function, which will reduce the effect from $\delta_{\eta_r}$ and $\delta {\rm Area}$ (see eq.~(\ref{eq:B_total_simple})), and to the approximation in the dependence of $\Theta^{(2)}$ on $\delta$Area. Still, we will see that it reproduces quite well  the results of the main calculation. 

For the perturbations due to recombination we are dealing with, as we saw in eq.~(\ref{guesstwo}),  $\delta_X$ runs over \{$\delta_\eta$ entering in $S^{(1)'}$\}, $\delta_{k_D}$, $\delta_{k_g}$ and $\delta$Area. Since later when we do the full calculation, we find that most of the signal is peaked on squeezed triangles with sides $l_1\sim 200$ and $l_2\sim l_3\sim 3000$, we concentrate our discussion directly on this limit. The spherical Bessel function $j_l$ in the integral in $k$ in each term of (\ref{blllsecond}) forces the $k$ of $\delta_X(k)$ to be approximately equal to $l/\eta_0$. This means that when a term has $l=l_1$, as  the first one shown in (\ref{blllsecond}), it represents an effect due to the low-$k$ part of the perturbation  $\delta_e$, while when it has $l=l_2$, as the second one shown in (\ref{blllsecond}), it represents an effect due to the high-$k$ part of the perturbation $\delta_e$. For  each of these separate terms, in Figures \ref{fig:btot}-\ref{fig:bH} we plot the approximate contributions to the signal-to-noise per triangle of the three-point function in the squeezed limit as obtained from (\ref{blllsecond}). The plots are for $l_1=200$. 

\begin{figure}
\begin{center}
\includegraphics[width=15cm]{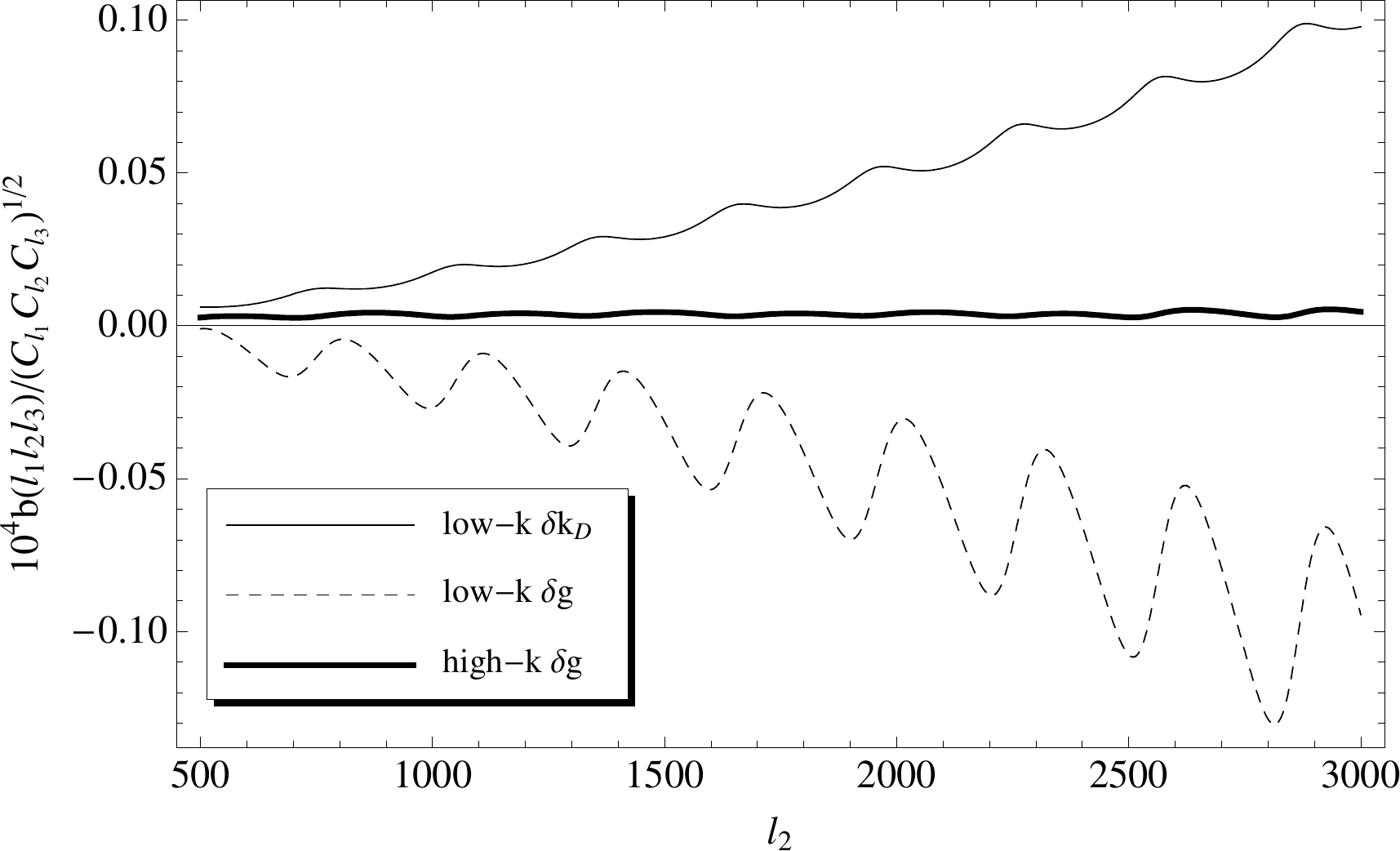}
\end{center}
\caption{\small \small  Approximate contributions to the signal-to-noise of the three
point function per triangle in the squeezed limit as obtained from
(\ref{blllsecond}) for $l_1=200$ as a function of $l_2=l_3$. The
contribution to the bispectrum from $\delta_{k_D}$ comes from the
perturbation due to $\delta n_e$. The different contributions to the
bispectrum from $\delta_g$ are shown in Figures \ref{fig:bL} and
\ref{fig:bH}. 
}\label{fig:btot}
\end{figure}

Let us start discussing the contribution arising from a low $k_e$ and high $k_s$. The contributions from \{$\delta k_D$ due to $\delta n_e$\} and $\delta g$ are shown in Fig.~\ref{fig:btot}. As discussed before, the low-$k_e$ contributions nearly cancel each other, which is reminiscent of the fact that for this low-$k_e$ modes recombination happens as in the unperturbed universe, just a bit time translated. This can be also confirmed by comparing the bispectra they generate: the effect from the low-$k_e$ $\delta_{k_g}$ is shown separately in Fig.~\ref{fig:bL}, and one can see that it is very nearly equal to minus the effect of the low-$k_e$ $\delta_{k_D}$ shown in Fig.~\ref{fig:btot}. Since $k_e$ is fixed and small, we have $k_s\sim l_2/\eta_0$. Notice also that from eq.~(\ref{guesstwo}) we could have expected that each of these two contributions should scale as $l_2^2\propto k_s^2$, which is nicely confirmed by these plots.

From Fig. \ref{fig:bL} we also confirm that the phase shift of the source due to $\delta\eta_r$ has a contribution to the bispectrum which increases linearly with $k_s$ and is non-negligible, as suggested by our analysis of its contribution to $\Theta_l^{(2)}$ (see Fig. \ref{fig:kDL}). In the full calculation this linear growth will become milder at higher $k_s$'s because there is a time integral which tends to cancel out the effect. 

\begin{figure}
\begin{center}
\includegraphics[width=15cm]{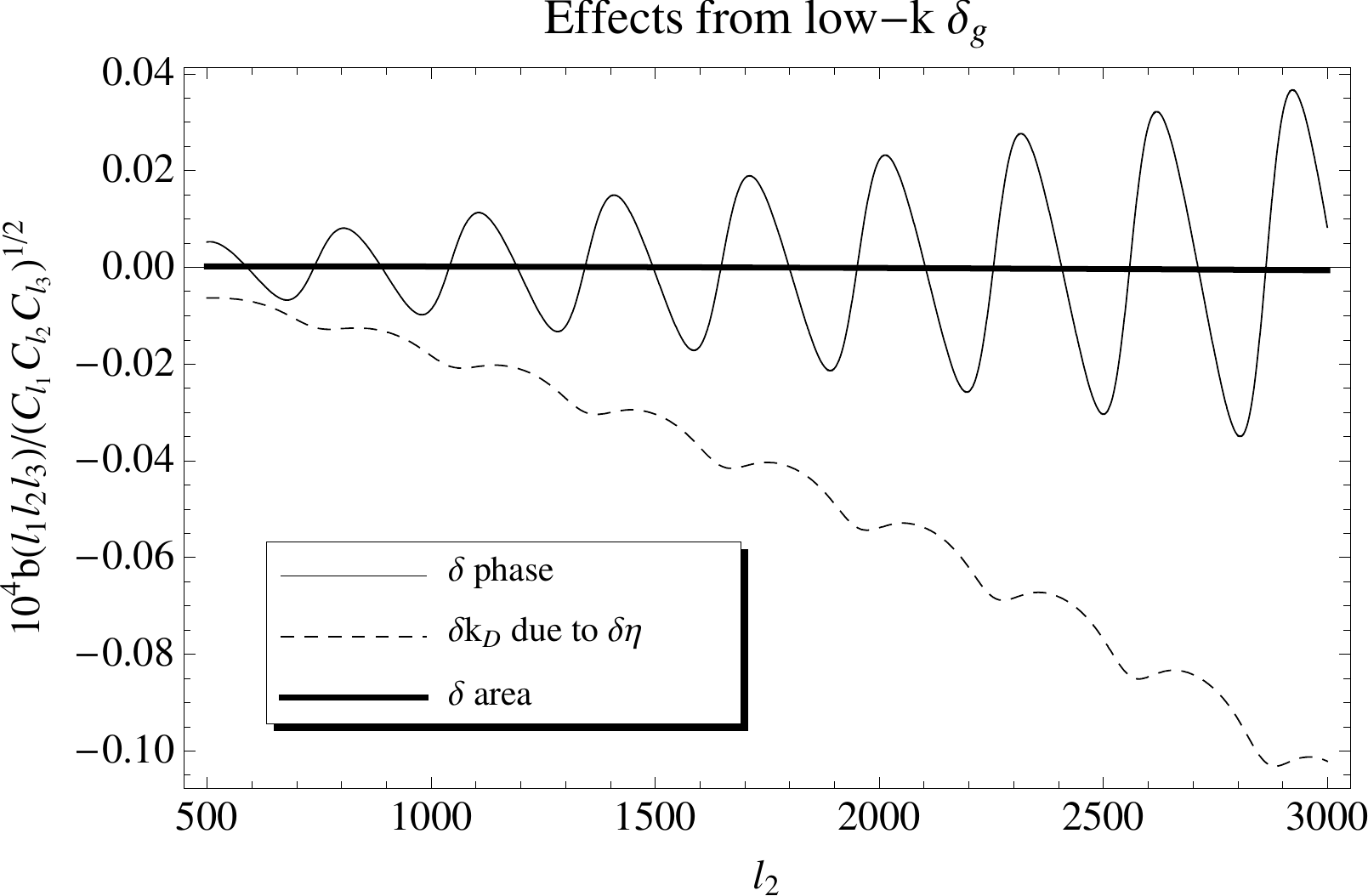}
\end{center}
\caption{\small \small  The different contributions (obtained from the approximation given in eq.~(\ref{blllsecond})) to the
signal-to-noise of the bispectrum per triangle due to a low-$k$
$\delta_g$  for $l_1=200$ as a function of $l_2=l_3$. The sum of all these curves is equal to the low-$k$ $\delta_g$ curve in Fig.~\ref{fig:btot}.  Due to the
ensemble averaging of the primordial fluctuations, we have $k_e\approx
l_1/\eta_0$, and $k_s\approx l_2/\eta_0$. The $\delta k_D$ due to
$\delta\eta$ is comparable and opposite in sign to the $\delta k_D$
due to $\delta n_e$ (see Fig.~\ref{fig:btot}). The scalings of the
bispectra due to the two contributions to $\delta {k_D}$ follows
$k_s^2\propto l_2^2$ as implied by eq.~(\ref{guesstwo}). The change in the
phase of the first order source due to the shift of the position of
the last scattering surface brings an oscillating contribution
proportional to $k_s$, whose amplitude grows linearly with $k_s$, as expected from eq.~(\ref{guesstwo}). 
}\label{fig:bL}
\end{figure}

Let us now consider the other interesting regime. The combination of a fixed low $k_s\eta_0=200$ with a high $k_e$ is shown in Fig. \ref{fig:bH}. As anticipated (see Fig.~\ref{fig:kDH}) we find that there are no important contributions in this regime.

\begin{figure}
\begin{center}
\includegraphics[width=15cm]{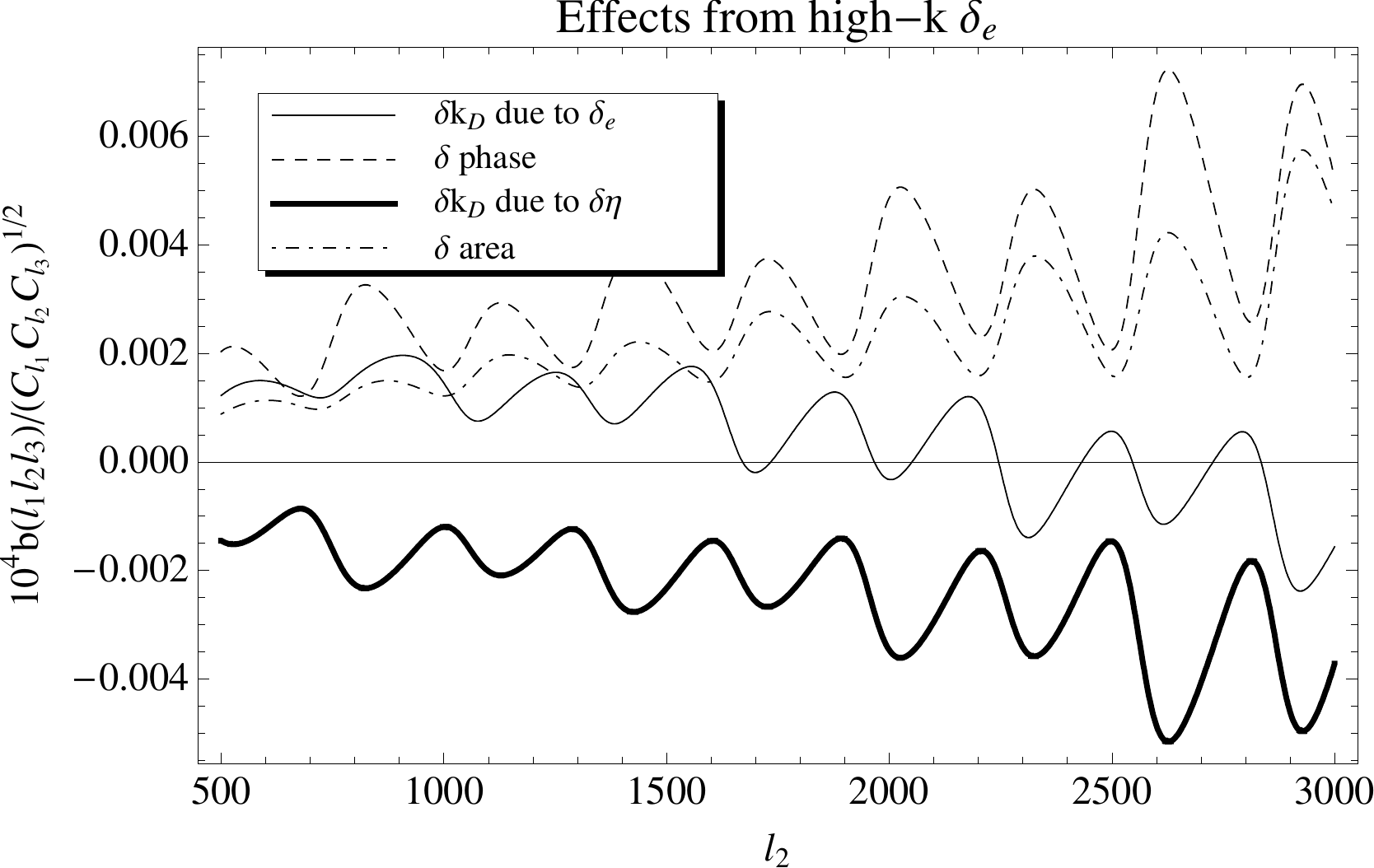}
\end{center}
\caption{\small \small  \small The different contributions (obtained
from the approximation given in eq.~(\ref{blllsecond})) to the
signal-to-noise of the bispectrum per triangle due to a high-$k$
$\delta_e$ for $l_1=200$ as a function of $l_2=l_3$. Here we have
$k_e\approx l_2/\eta_0$, and $k_s\approx l_1/\eta_0$. Although a high-$k_e$ perturbation of order 1 can produce a large bispectrum (cf. eq.~(\ref{highkorderone})), all high $k_e$ effects due to recombination are suppressed and therefore, we expect that the bispectrum generated in this regime to be small.
}\label{fig:bH}
\end{figure}


Let us now explain the signs of the contributions proportional to $\delta k_D$. We can concentrate on a low-$k_e$ perturbation since this is the regime in which they are relevant. In this limit, the most important terms of the bispectrum in (\ref{blllsecond}) are just the ones where $k_e\sim l_1/\eta_0$, which we can now schematically write as $\Theta_{l_1}\times\delta\left|\Theta_{l_2}\right|^2$. This  tells us how the amplitude of the high-frequency modulations changes in the presence of a long-wavelength mode. If the bispectrum is positive,  high-$l$ modes have an enhanced amplitude in hotter long-wavelength regions of the sky, while they have a smaller amplitude in colder long-wavelength regions~\footnote{This means that a positive bispectrum implies a positive skewness of the one-point distribution function of the CMB temperature anisotropies.}. 

With this intuition in mind, let us now check whether the different mechanisms generating $\Theta_l^{(2)}$ produce bispectra with the signs obtained with our approximate equation (\ref{blllsecond}). 
Let us concentrate on the interesting regime, corresponding to the low-$k_e$ (correspondoing to low $l_1$) $\delta_{k_g}$ terms. Notice that, since $\Theta_{l_1}(\eta_0)\approx S^{(1)}(k=l_1/\eta_0,\eta_*)$, a low-$l$ mode at the scale of the first acoustic peak corresponds to a positive temperature perturbation on the sky, which is also associated  to overdensities. But as we discussed in our former paper \cite{inprep1}, in overdensities $n_e$ is larger and thus $\delta_{\eta_r}$ is positive around the same scale (as can be seen by looking at the phase shift term in Fig. \ref{fig:kDL}). Therefore, in the presence of such a low-$l$ mode, the photons have more time to diffuse, which decreases the amplitude of the high-$l_2$ $\Theta_{l_2}$. This means that we can expect $\delta_{\eta_r}$ entering in $k_D$ (i.e. the $\delta_{k_g}$ effect) to generate a negative bispectrum, confirming what we find in Fig. \ref{fig:bL}. This conclusion holds even for higher $l_1$ (around the second acoustic peak), since there $\delta_{\eta_r}$ and $\Theta_{l_1}^{(1)}$ have again the same signs. 

In the case of the  $\delta_{k_D}$ perturbation  generated by $\delta n_e$, the perturbation is clearly positive in low-$k$ overdensities. By exactly following the same argument we made for $\delta_{k_g}$, it is immediate to see that the induced bispectrum in this case is positive, confirming what found in Fig.~\ref{fig:btot}.

From these estimates we see that the bispectrum produced by perturbations to the recombination history can be quite large, though some cancellations tend to reduce its net size. This clearly deserves a more accurate computation, which we are going to do in the remaining of this paper. In order to do this, we will work well inside the horizon, so that we expect to be able to neglect second order metric perturbations; and we will also drop, in the second order photon Boltzmann equation, all those quadratic second order terms that are not proportional to $\delta_e$.  This procedure will grasp the leading effects proportional to $\delta_e$ which are expected to be enhanced.

\section{Second order temperature anisotropies\label{sec:second_order}}
\subsection{Line of sight integral}

Let us procede to find the exact expression corresponding to our intuitive expression (\ref{guess}). 
We follow the conventions of \cite{inprep1}. As explained, we work at first order in the metric perturbations, restricting our calculation to be valid for scales much shorter than the horizon at recombination. In this analytical part we work in Newtonian gauge. The line 
element is given by:
\begin{eqnarray}\label{metric}
ds^2=a^2(\eta )\left[-(1+2\Psi) d\eta ^2+(1-2\Phi)\delta_{ij}dx^idx^j\right]\ ,
\end{eqnarray}
where we ignore primordial vector or tensor modes.

We start with the second order brightness equation given by  eq.~(76) in \cite{inprep1}. It is an 
equation for the second order perturbations $\Delta^{(2)}$ of the photon energy density:
\begin{eqnarray}\label{delta}
\Delta^{(i)}(\vec{x},\vec{n},\eta)\equiv\frac{\int dp\, p^3 F^{(i)}}{\int dp\, p^3 F^{(0)}}\ ,
\end{eqnarray}
for $i=1,2$ , where $$F(\vec x,p^i,\eta)=F^{(0)}(\vec x,p^i,\eta)+F^{(1)}(\vec x,p^i,\eta)+\frac{1}{2}F^{(2)}(\vec x,p^i,\eta)\ ,$$ is the gauge invariant photon one-particle distribution function written to second order.  $x^i$ are comoving coordinates. The photon momentum is given by $p^i=pn^i$ with $n^2=1$. The $p^i$'s are the momenta defined in the inertial frame locally defined at rest at the point $x^i$ (see \cite{inprep1} for details). This choice of $\Delta$ in useful as it integrates out the momentum dependence from the photon one-particle distribution. 

In analogy with the first order calculation \cite{cmbfast}, the full (first + second order) brightness equation \cite{inprep1}  can be written as:
\begin{eqnarray}\label{los}
&&e^{\int^\eta_0 d\eta' \dot\tau(\eta') (1+\delta_e(\vec{x}(\eta'),\eta'))} \frac{D}{D\eta}\left[
\left(\Delta^{(1)}+\frac{1}{2}\Delta^{(2)}+4\Psi\right)e^{-\int^\eta_0 d\eta' \dot\tau(\eta') 
(1+\delta_e(\vec{x}(\eta'),\eta'))}\right]= \nonumber \\ \nonumber 
&&=-\dot\tau \left[\Delta^{(1)}_0+4\Psi+\frac{1}{2}\Delta^{(2)}_{00}
-\frac{1}{2}\sum_{m=-2}^2 \sqrt{\frac{4\pi}{5^3}}\left(\Delta^{(1)}_{2m}+\frac{1}{2}\Delta^{(2)}_{2m}
\right)Y_{2m}(\vec{n})+
4 \vec{n} \cdot (\vec{v}^{(1)}+\frac{1}{2}\vec{v}^{(2)})\right. \\  
 &&\left.
+\delta_e(x(\eta))\left(\Delta_0^{(1)}+4\Psi-\frac{1}{2}\sum_{m=-2}^2 \sqrt{\frac{4\pi}{5^3}}
\Delta^{(1)}_{2m}Y_{2m}(\vec{ n})+4 \vec{v}^{(1)} \cdot \vec{n}\right) \right]\ ,
\end{eqnarray}
where $\dot\tau=-n_e\sigma_T a$ is the differential optical depth, $\sigma_T$ is the Thomson cross 
section, $a$ is the scale factor. A dot denotes a derivative with respect to conformal time. In the above $\Delta^{(i)}=\Delta^{(i)}(\vec{x}(\eta),\hat n,\eta)$, $\vec v=\vec{v}(\vec{x},\eta)$ and $\Psi=\Psi(\vec{x},\eta)$. $\vec{v}$ is the  baryon fluid velocity; $Y_{lm}$ are the spherical harmonics
~\footnote{\label{footnote}Two technical notes. 

First, in the brightness equation in real space, we are not using the Legendre 
functions $P_l(\hat{{v}}_b\cdot\vec{n})$ for the first order terms entering the second order brightness 
equation as written in both \cite{astro-ph/0703496} and \cite{bartolo}. Using the Legendre Polinomials is only justified for the  first order terms in Fourier space when the choice for the $z$ axis, $\hat{ e}_z$, is along the axis of  symmetry, i.e. along $\vec{k}$ of the perturbation for each mode. However, at second order, one cannot choose a common axis of symmetry 
for the two Fourier modes in the convolution, and so in real space the expression is more complicated (though it simplifies in Fourier space). This is why in our expression we are ending up with a multipole expansion in both $l$ and $m$ using 
spherical harmonics.

Second, in solving the brightness equation, we are taking the approach of solving the full (first + second order) brightness equation, and {\it then} expanding the 
line of sight solution to second order. This approach gives a result where the piece of the source multiplying the $\delta_e$-containing terms in eq.~(\ref{guesstwo}) is not tight coupling suppressed, and has exactly the same structure as the first order result (\ref{firstorder}) that allows us for a  straightforward interpretation of the results. 
}. Our convention for the spherical harmonic decomposition is:
\be
\Delta_{lm}=(-i)^{-l}\sqrt{\frac{2l+1}{4\pi}}\int d\Omega \Delta(\hat n)Y^\star_{lm}(\hat n)\ .
\ee

In writing eq.~(\ref{los}) we have already dropped all second order metric perturbations together with all products of first 
order terms not enhanced by $\delta_e$ (the brightness equation is given in full by eq.~(76) in \cite{inprep1}), and the integrated Sachs-Wolf effect. Our approach is in fact to concentrate on the terms enhanced by $\delta_e$  and to neglect metric perturbations at second order.  This last point will mean that our calculation will be valid only for multipoles well within the horizon at recombination. 

Due to the advective derivative in the streaming part of the Boltzmann equation, the streaming terms are 
evaluated along photon trajectories, i.e. along the line of sight. Since we neglect second order metric 
perturbations and first order metric perturbations not multiplied by $\delta_e$, photon trajectories are straight and the position $\vec{x}$ along the line of sight is a 
function of $\eta$ through
\begin{equation}\label{xeta}
\vec{x}(\eta)=\vec{x}_0+\hat{{n}} \eta \ .
\end{equation}
 In doing this, we neglect the effects of gravitational lensing which does not contribute significantly to the part of the 3-point function generated around recombination.

Next, we can perform the line of sight integration as it is done at first order \cite{cmbfast} using the above $x$ dependence on $\eta$.  We then take the second order piece of the result, and Fourier transform it to find:
\begin{eqnarray}\label{bright2}
&&\Delta^{(2)}(\eta_0,\vec{k},\hat n)=2 \int_0^{\eta_0} d\eta\, g^{(0)}(\eta)S(\vec{k},\eta,\hat n)\ .
\end{eqnarray}
The visibility function is $g(\eta)=-\dot\tau e^{-\tau}$, and the second order source function is given by
\begin{eqnarray}\label{source-1}  
&&S(\vec{k},\eta,\hat n)= e^{-i \vec{k} \cdot \vec{n}
(\eta_0-\eta)}\left\{\frac{1}{2}\Delta_{00}^{(2)}(\eta,\vec{k})-\frac{1}{4}\sum_{m=-2}^2 \sqrt{\frac{4\pi}{5^3}}
\Delta^{(2)}_{2m}(\eta,\vec{k})Y_{2m}(\vec{n})+2 \vec{n}\cdot\vec{v}^{(2)}(\vec{k},\eta)+ \right. \nonumber\\
\nonumber 
&& + \int \frac{d^3\tilde k}{(2\pi)^3} \left(\delta_e(\vec{k}-\vec{\tilde{ k}},\eta)-\int^\eta_{\eta_0}
d\eta' 
e^{i (\vec{k}-\vec{\tilde k}) \cdot \vec{n}(\eta'-\eta)}
\dot\tau(\eta') \delta_e(\eta',\vec{k}-\vec{\tilde{ k}})\right)\nonumber \\
&&\left. \times\left(
\Delta_0^{(1)}(\vec{\tilde{k}})+4\Psi(\vec{\tilde{k}})+\frac{1}{2}\Delta_2^{(1)}(\vec{\tilde{k}}) 
P_2(\hat{\vec{\tilde{k}}} \cdot \vec{n})+4 \vec{v}^{(1)}(\vec{\tilde{k}}) \cdot \vec{n}\right)\right\} \ .
\end{eqnarray}
The extra phase in the line-of-sight integral over $\delta_e$ above comes from the dependence of $\delta_e$ on $\vec{x}(\eta')$, and not on $\vec{x}(\eta)$ in (\ref{los})~\footnote{Ref. \cite{Khatri:2009ja} has correctly pointed out the presence of an algebraic mistake in this point in the first version of this paper. After accounting for the mistake, our numerical results are marginally affected by this.}.

To obtain $\Delta^{(2)}(\vec{ x}=0,\eta_0)$, for an observer today at the origin, we integrate (\ref{bright2}) 
over $d^3k/(2\pi)^3$. 
We then expand the second order anisotropies in multipoles to obtain:
\begin{eqnarray}\label{Delta2lm1}  
\Delta^{(2)}_{lm}(\eta_0, \vec{x}=0)&=&2\int \frac{d^3 k}{(2\pi)^3} \int^{\eta_0}_0 d\eta g^{(0)}(\eta) (-i)^{-l}\sqrt{\frac{2 l+1}{4\pi}}\int d\hat n\,Y_{lm}^*(\hat n)S(\hat n,\vec k,\eta)\ ,
\end{eqnarray}

The source can be split into two pieces as follows:
\be
S=S^{a}+S^{b}\ ,\label{sourcelm} 
\ee
where
\be
&&S^{a}\equiv e^{-i \vec{k} \cdot \vec{n}
(\eta_0-\eta)}\sum_{\tilde l,\tilde m} (-i)^{\tilde l}\sqrt{\frac{4\pi}{2\tilde l+1}}Y_{\tilde l,\tilde m}(\hat n)\left\{
\frac{1}{2}\Delta_{00}^{(2)}(\eta,\vec{k})\delta_{\tilde l,0} \delta_{\tilde m,0}+\frac{1}{20}
\Delta^{(2)}_{2\tilde m}(\vec{k})\delta_{\tilde l,2}\right.\nonumber \\     
&&  \left.\quad\quad+2 \delta_{\tilde l,1}\left[i \delta_{\tilde m,0} v_z^{(2)}(\vec{k})+\frac{i}{\sqrt{2}}v^{(2)}_x(\vec{k})(-\delta_{\tilde m,
1}+
\delta_{\tilde m,-1})-\frac{1}{\sqrt{2}}v^{(2)}_y(\vec{k})(\delta_{\tilde m,1}+\delta_{\tilde m,-1})\right]\right\} \ , \label{Sa}\\ 
&&S^{b}=\sum_{\tilde l,\tilde m}(-i)^{\tilde l}  \sqrt{
\frac{4 \pi}{2 \tilde l+1}}Y_{\tilde l \tilde m}(\hat n)\int \frac{d^3\tilde k}{(2\pi)^3} \tilde
S^{b}_{\tilde l}(|\vec{k}-\vec{\tilde k}|,|\vec{\tilde k}|,\eta) 
 Y^*_{\tilde l\tilde m}(\hat{\tilde{k}})\times \nonumber \\
&&\ \ \ \ \ \ \times e^{-i \vec{\tilde k} \cdot \vec{n}
(\eta_0-\eta)}e^{-i (\vec{k}-\vec{\tilde k}) \cdot \vec{n}
(\eta_0-\tilde\eta)}
\zeta(\vec{k}-\vec{\tilde k})\zeta(\vec {\tilde k})\ ,\label{eq:source_redefinition}
\\ \nonumber 
\ee
with
\be
\tilde S^{b}_l(|\vec{k}-\vec{\tilde k}|,|\vec{\tilde k}|,\eta)=
\tilde
S^{b,1}(|\vec{k}-\vec{\tilde k}|,\eta)\tilde S^{b,2}_l(|\vec{\tilde k}|,\eta)\ ,
\ee
where
\be  \label{eq:sbdetails}
&&\tilde
S^{b,1}(|\vec{k}-\vec{\tilde k}|,\eta)\equiv\delta_e(|\vec{k}-\vec{\tilde{k}}|,\eta)-\int^\eta_{\eta_0}*
d\eta' \dot\tau(\eta') \delta_e(|\vec{k}-\vec{\tilde{ k}}|,\eta')\ ,\\    
&&\tilde S^{b,2}_l(|\vec{\tilde k}|,\eta)\equiv
\left(\Delta_0^{(1)}(\tilde{{k}})+4\Psi(\tilde{{k}})\right)\delta_{l,0}\delta_{m,0}\sqrt{4\pi}+4
\sqrt{\frac{4\pi}{3}}v_0^{(1)}(\tilde{{ k}}) \delta_{l,1}-
\frac{1}{2}\sqrt{\frac{4\pi}{5}}\Delta^{(1)}_2(\tilde{{k}})\delta_{l,2} \label{Sb}\ .
\ee
In the definition of $S^b$ we have already extracted the primordial density perturbation $\zeta$. The fluctuations in $\tilde S^{b,i}$ are to be meant as just the first order transfer functions. $\tilde S^{b,2}$ is nothing but the first order source. 
Also we defined $\tilde\eta=\eta$ for all pieces of the source (\ref{source-1}), except the piece containing the line-of-sight integral of $\delta_e$ over $\eta'$, for which $\tilde\eta=\eta'$. For that piece, the exponent $e^{-i (\vec{k}-\vec{\tilde k}) \cdot \vec{n}
(\eta_0-\tilde\eta)}$ must be moved inside the integral over $\eta'$ contained in $\tilde S^b_{l}$, which we do not do explicitly for clarity. To remind ourselves about this, we put a star after the integral sign over $\eta'$ above.

The splitting of the source between the $(a)$ and $(b)$ term is quite physical. Part $(a)$ of the source contains the second order monopole, dipole and quadrupole, which takes into account, as we will shortly see,  of the 
perturbation to the photon diffusion scale due to $\delta n_e$. Part $(b)$ of the source contains products of first order terms, 
and, as we will see, it corresponds to the perturbation of the visibility function. 

\subsection{Perturbing the diffusion length: part $(a)$ of the source \label{sec: part a}}

Unlike $S^{b}$ which is the convolution of two first-order pieces, part $(a)$ of the source (\ref{sourcelm}) 
contains the lowest moments of the second order photon energy density perturbations. One should solve 
the second order Boltzmann hierarchy to obtain those exactly. In this section we will find an approximate solution for the first  second-order multipoles, where we are able to write them as a convolution of two first order perturbations. The result for the source $S^{a}$ will be of the same form 
as $S^{b}$ in (\ref{eq:source_redefinition}), and below we will derive the analogous expressions for $\tilde S^{a,1}$ and $\tilde 
S_l^{a,2}$.

Here we are interested in the effect of $\delta_e$ on the first second-order multipoles. It is easy to realize that this effect is tight coupling suppressed. This does not mean that the effect is negligible: even a relatively small alteration of the Silk damping scale can lead, as we will see, to rather large effects. For this reason, the effect will be large for those high-$k$ modes for which the Silk damping is relevant ($k\gtrsim k_D\simeq 0.15\ $Mpc$^{-1}$). Further, as we  have shown in \cite{inprep1}, at high $k$'s the perturbation to $\delta_e$ gets damped. This is due to the fact that if the mode oscillates too fast with respect to the timescales of recombination, the perturbation to the free electron density averages out to zero. For these two reasons, we can restrict our interest to the second order perturbation generated on the monopole, dipole and quadrupole at very high $k$'s as sourced by a relatively long wavelength $\delta_e$. In this limit, a good approximation to our 
solution can be
found by just perturbing the diffusion damped solutions for the first order moments, given in 
e.g. \cite{Dodelson}, eq.s~(8.39, 8.40). If we imagine to have a very long wavelength mode $\delta_e$, its main effect on the evolution of short scale perturbations will be to alter the local density of free electrons. We can take this into account by substituting
\begin{equation}
n_e(\eta)\rightarrow n_e(\eta) (1+\delta_e(x,\eta))
\end{equation}
in eq.~(8.40) of \cite{Dodelson} for the damping scale.  We can therefore write, for $l=0,1,2$:
\begin{equation}\label{toperturbR}  
\Delta_l^{(1)+(2)}(\vec k,\eta)\simeq\Delta_l^{(1)}(\vec k,\eta)\left(1-k^2\int_0^{\eta}d\eta'
\frac{1}{6(1+R)\dot\tau(\eta')}\left[\frac{R^2(\eta')}{(1+R)}+\frac{8}{9}\right]\delta_e(x,\eta')
\right)\ ,
\end{equation}
where $R\equiv 3\rho^{(0)}_b/(4\rho^{(0)}_\gamma)$.  This is a good approximation in the limit where the wavelength of the temperature perturbation is much shorter than the one of the $\delta_e$ perturbation. Using this equation at second order amounts also to neglecting the possible azimuthal dependence of $\Delta^{(2)}(k)$, which again should be a good approximation in the same limit. Notice also that we did not need to perturb $R$ in (\ref{toperturbR}), since it would give rise to a second order perturbation not proportional to $\delta_e$.


For modes with no azimuthal dependence, the relation between Legendre moments and spherical harmonic
moments is (with $k$ not necessarely on the $z$ axis):
\begin{equation}\label{eq: legendre_to_spherical}
\Delta_{lm}(\vec{k},\eta)=\sqrt{(4\pi)(2l+1)}(-i)^{-l}\Delta_l(\vec{k},\eta )Y^*_{lm}(\hat{{ k}}) \ .
\end{equation}
Using the above equation also at second order, we can write:
\begin{equation}
\Delta_{lm}^{(1)+(2)}(\vec k,\eta)=\Delta_{lm}^{(1)}(\vec k,\eta)\left(1-k^2\int_0^{\eta}d\eta'
{\rm ker}(\eta')\delta_e(x,\eta') \right)\ , 
\end{equation}
where we have defined
\begin{equation}
{\rm
ker}(\eta')=\frac{1}{6\left(1+R(\eta')\right)\dot\tau(\eta')}
\left[\frac{R(\eta')^2}{(1+R(\eta'))}+\frac{8}{9}\right]\ .
\end{equation}
In real space, this equation reads
\begin{equation}
\Delta_{lm}^{(1)+(2)}(\vec{x},\eta)=\Delta_{lm}^{(1)}(\vec{x},\eta)+\nabla^2\Delta_{lm}^{(1)}(\vec{x},\eta)
\int_0^{\eta}d\eta' {\rm ker}(\eta')\delta_e(\vec{x},\eta')\ .
\end{equation}
Now, we can take into account of the mild dependence of $\delta_e$ on the position by Fourier transforming properly the above to get
\begin{equation}  
\Delta_{lm}^{(1)+(2)}(\vec{k},\eta)=\Delta_{lm}^{(1)}(\vec{k},\eta)-\int
\frac{d^3 k'}{(2\pi)^3}\int_0^{\eta}d\eta' {\rm
ker}(\eta')\delta_e(\vec{k}-\vec{ k}',\eta')
{k'}^2\Delta_{lm}^{(1)}(\vec{ k}',\eta)\ .
\end{equation}
We have therefore obtained an expression for $\Delta_{lm}^{(2)}$ which is valid in the `squeezed' limit where the space dependence of $\delta_e$ is milder than the one of $\Delta^{(1)}_{lm}$:
\be\label{Deltalm2}
\Delta_{lm}^{(2)}(\vec{k},\eta) & =  & -2\int \frac{d^3
k'}{(2\pi)^3}\int_0^{\eta}d\eta' {\rm
ker}(\eta')\delta_e(\vec{k}-\vec{{ k}}',\eta')
{k'}^2(-i)^{-l}\nonumber \\
   &&\times \sqrt{(4\pi)(2l+1)}\Delta_{l}^{(1)}(\vec{ k}',\eta)Y^*_{lm}(\vec{ k}')
\ee
with $l=0,1,2$.

Part $(a)$ of the source (\ref{sourcelm}) can be written as (for $\vec{k}$ along $\hat{{z}}$):
\be  
S^{(a)} & = & e^{-i \vec{k} \cdot \vec{n}
(\eta_0-\eta)}\sum_{\tilde l,\tilde m} (-i)^{\tilde l}\sqrt{\frac{4\pi}{2\tilde l+1}}Y_{\tilde l,\tilde m}(\hat n)\left\{ \frac{1}{2}\Delta_{00}^{(2)}(\vec{k},\eta)\delta_{\tilde l,0}\delta_{\tilde m,0}+\right.\nonumber\\
& & +\left.2\delta_{\tilde l,1}\left(
v_0^{(2)}(\vec{k})\delta_{\tilde m,0}-v_1^{(2)}(\vec{k})\delta_{\tilde m,1}-v_{-1}^{(2)}(\vec{k})\delta_{\tilde m,-1}\right)+\frac{1}{20}\delta_{\tilde l,2}\Delta_{2\tilde m}^{(2)}\right\}\ ,
\ee
where, for $\hat k=\hat z$, $v_m^{(2)}$ is defined by:
\be
v^{(2)}(\vec k)=-i v_0^{(2)}(\vec k) \hat z+\sum_{m=\pm 1} v^{(2)}_m(\vec k)\frac{\hat x\mp \hat y}{\sqrt{2}}\ .
\ee
In tight coupling we have the usual relationship
\begin{equation}
v^{(2)}_m(\vec{k})=(-1)^m\frac{\Delta_{1m}^{(2)}(\vec{k})}{4}\ ,
\end{equation}
so that we can write:
\begin{equation}
S^{a}(\vec{k})=e^{-i \vec{k} \cdot \vec{n}
(\eta_0-\eta)}\sum_{\tilde l,\tilde m} (-i)^{\tilde l}\sqrt{\frac{4\pi}{2\tilde l+1}}Y_{\tilde l,\tilde m}(\hat n)\left\{ \frac{1}{2}\Delta_{00}(\vec{k},\eta)\delta_{\tilde l,0}\delta_{\tilde m,0}+\frac{1}{2}\delta_{\tilde l,1}\Delta_{1,\tilde m}^{(2)}(\vec{k},\eta)+\frac{1}{20}\delta_{\tilde l,2}\Delta_{2\tilde m}^{(2)} \right\}\ .
\end{equation}
Though this formula is written now for generic $\vec k$, we have derived it using formulas valid only for $\vec k$ along $\hat z$. By using the transformation rules under rotations that we give in App.~\ref{appendix:A}, one can show  that this expression transforms correctly under rotations, and is correct for the frame where $\vec k$ is parallel to $\hat z$. Therefore it is correct in any frame.
Combining this with (\ref{Deltalm2}) we obtain:
\be\label{SALM}
S^{a}&=&\sum_{\tilde l,\tilde m}i^{-\tilde l}  \sqrt{
\frac{4 \pi}{2 \tilde l+1}}Y_{\tilde l \tilde m}(\hat n)\int \frac{d^3\tilde k}{(2\pi)^3} \tilde
S^{a}_{\tilde l}(|\vec{k}-\vec{\tilde k}|,|\vec{\tilde k}|,\eta) 
 Y^*_{\tilde l\tilde m}(\hat{\tilde{k}})\times \nonumber \\
&&\ \ \ \ \ \ \times e^{-i \vec{\tilde k} \cdot \vec{n}
(\eta_0-\eta)}e^{-i (\vec{k}-\vec{\tilde k}) \cdot \vec{n}
(\eta_0-\tilde\eta)}
\zeta(\vec{k}-\vec{\tilde k})\zeta(\vec {\tilde k})\ ,
\ee
where we kept $\tilde\eta$ (although it equals $\eta$ in this case), which will simplify the calculation of the bispecturm later on. In the above equation we defined:
\be  
&&\tilde S^{a}_l(|\vec{k}-\vec{ k}'|,|\vec{ k}'|,\eta)=
\tilde
S^{a,1}(|\vec{k}-\vec{ k}'|,\eta)\tilde S^{a,2}_l(|\vec{ k}'|,\eta)\ ,
\ee
with
\be
&&\tilde
S^{a,1}(|\vec{k}-\vec{ k}'|,\eta)\equiv -\int_0^{\eta}d\eta' \delta_e(|\vec{k}-\vec{ k}'|,\eta')
{\rm ker}(\eta')\ ,  \\  
&&\tilde S^{a,2}_l(|\vec{ k}'|,\eta)\equiv
\sqrt{4\pi} {k'}^2\Delta_0^{(1)}({{k}}',\eta)\delta_{l,0}+4\sqrt{\frac{4\pi}{3}}{ k'}^2
v^{(1)}({{k}}',\eta)\delta_{l,1}-\frac{1}{2}\sqrt{\frac{4\pi}{5}}\Delta_2^{(1)}(k',\eta) k'^2\delta_{l,2}\nonumber\ .
\ee
Thus, as anticipated, we find that even Part ($a$) of the source can be written in  exactly the same form as in eq.~(\ref{eq:source_redefinition}). As we already did in the case of $S^{(b)}$, in $S^{(a)}$ we have introduced the primordial fluctuation $\zeta$.  The first order quantities in $\tilde S^{a,i}$ should be meant simply as transfer functions. Notice that 
\be 
\tilde S^{a,2}\sim k_D^2 \frac{\partial S^{(1)}}{\partial \log k_D}\ ,
\ee 
where $S^{(1)}$ is the first order source function.

\subsection{Line of sight solution}

Now that we have calculated the second order source, we can put $\Delta_{lm}^{(2)}$ in a form that is keen for the calculation of the 3-point function. We substitute the expressions for the source, (\ref{eq:source_redefinition}) and (\ref{SALM}), into eq.~(\ref{Delta2lm1}) and expand the exponentials into spherical harmonics. Performing the $\hat n$ integral, we obtain:
\begin{eqnarray}\label{deltalm2main}  
&&\Delta^{i,(2)}_{lm}(\eta_0, \vec{x}=0)=\\ \nonumber 
&&
\sum_{l',l'',\tilde l, L}\sum_{m',m'',\tilde m,M} 2 \int \frac{d^3k}{(2\pi)^3}\frac{d^3\tilde k}{(2\pi)^3}
\int^{\eta_0}_0 d\eta g(\eta)i^{l-l'-l''-\tilde l}4\pi(2 l+1)(2 L+1)\sqrt{(2 l'+1)(2 l''+1)}\\ \nonumber
&&\times j_{l'}(\tilde k(\eta_0-\eta))j_{l''}(|\vec k-\vec{\tilde k}|(\eta_0-\tilde\eta)) Y_{l'm'}^*(\hat{\tilde{k}})Y_{l''m''}(\vec{k}-\vec{\tilde{k}}) Y_{\tilde l\tilde m}^*(\hat{\tilde{k}})
S^{i,1}(|\vec{k}-\vec{\tilde{k}}|,\eta)S^{i,2}_{\tilde l}(\tilde{k},\eta)  \times \\ \nonumber 
&&\times
 \left(
\begin{array}{ccc}
l & l'' & L \\
0 & 0 & 0
\end{array}
\right)
\left(
\begin{array}{ccc}
l & l'' & L \\
-m & -m'' & -M
\end{array}
\right)
 \left(
\begin{array}{ccc}
L & l' & \tilde l \\
0 & 0 & 0
\end{array}
\right)
\left(
\begin{array}{ccc}
L & l' & \tilde l \\
M & m' & \tilde m
\end{array}
\right)\zeta(\vec k-\vec{\tilde k})\zeta(\vec{\tilde k})
\ ,
\end{eqnarray}
where $i$ goes over $a$ and $b$, and we have introduced the Wigner 3$j$ symbols. The above result is in a convenient form for calculating the 3-point function.

Before proceeding, we need to write down the first order perturbation in a similar manner. By analogy with the above calculation, the first order perturbation is given by:
\begin{equation}\label{delta1}
\Delta^{(1)}_{lm}(\eta_0,\vec{x}=0)=(-1)^l\int
\frac{d^3k}{(2\pi)^3}\Delta_l^{T,(1)}(k,\eta_0)Y_{lm}^*(\hat{{
k}})\zeta(\vec{k})\ ,
\end{equation}
where $\Delta_l^{T,(1)}(k,\eta)$ is the first order transfer function, that depends only on the
magnitude of $\vec{k}$:
\begin{eqnarray}\label{firstT}  
\Delta_l^{T,(1)}(k,\eta_0)&=&(-1)^l\sqrt{4\pi(2l+1)}\int^{\eta_0}_0
d\eta\; j_l(k(\eta_0-\eta)) \times \\ \nonumber
 &&
\left[\left(-\dot\tau e^{-\tau}\right)
\left(\Delta^{(1)}_0+4\Psi-\frac{1}{4}\Delta^{(1)}_2\right)-\frac{1}{i
k} \frac{d}{d\eta}\left(\left(-\dot\tau e^{-\tau}\right)\left( -4i
v_0^{(1)}\right)\right)-\right.\nonumber\\
&&\left.
-\frac{1}{k^2}\frac{d^2}{d\eta^2}\left(\left(-
\dot\tau e^{-\tau}\right)\frac{3}{4}\Delta^{(1)}_2 \right)\right]\nonumber\ .
\end{eqnarray}

\section{3-point function \label{sec: 3-point function}}
We are now ready to compute the 3-point function:
\begin{eqnarray}\label{3pt}
&&\langle\Delta_{l_1m_1}\Delta_{l_2m_2}\Delta_{l_3m_3} \rangle=\frac{1}{2}
\langle\Delta^{(2)}_{l_1m_1}
\Delta^{(1)}_{l_2m_2}\Delta^{(1)}_{l_3m_3}\rangle+ \  {\rm 2\ symmetric\ terms} \ .
\end{eqnarray}
Here $\Delta^{(1)}_{lm}$ is given by (\ref{delta1}) and $\Delta^{(2)}_{lm}$ is given by (\ref{deltalm2main}).
The main ingredients in the calculation of the above expectation value are the following. We assume gaussian primordial perturbations with a power spectrum $P(k)$ defined as
\begin{equation}\label{deltaf}
\langle\zeta(\vec{k}_1)\zeta(\vec{k}_2)\rangle=(2\pi)^3\delta^{(3)}(\vec{k}_1+\vec{k}_2)P(k_1)\ .
\end{equation}
Using the above expectation value in eq.~(\ref{3pt}), one obtains a term multiplied by $\delta^{(3)}(\vec{k}_1+\vec{k}_2+\vec{k}_3)$.
 The momentum dependence of $\tilde S^{i}$ (with $i=a,b$), is such that
one can do the integral in $\vec k_1$ trivially, as well as the two angular integrals on $\hat{{ k}}_{2,3}$. 

Using expression (\ref{eq: legendre_to_spherical}) for $\Delta_{lm}^{(1)}$, which is exact at first order, the final result of the above manipulations for the three point function arising from the source $S^i$ is
\begin{eqnarray}\label{3pttemp}  
&&\frac{1}{2}
\langle\Delta^{(2),i}_{l_1m_1}
\Delta^{(1)}_{l_2m_2}\Delta^{(1)}_{l_3m_3}\rangle= 
\sum_{L,l',\tilde l}\int \frac{k_2^2dk_2}{(2\pi)^3}\int \frac{k_3^2dk_3}{(2\pi)^3}\int^{\eta_0}_0 d\eta\;
g(\eta)\\
   \nonumber &&
 i^{l_1-l'-l_2-\tilde l} \sqrt{4\pi(2l_1+1)(2l_2+1)(2l_3+1)}\nonumber
\\ 
\nonumber
&& (2L+1)(2 l'+1)\sqrt{(2\tilde l+1)(2l_1+1)} j_{l'}(k_3(\eta_0-\eta))j_{l_2}(k_2(\eta_0-\tilde\eta))\\ \nonumber
&& \times \tilde
S^{i,1}(k_2)S^{i,2}_{\tilde l}(k_3)
\Delta^{T,(1)}_{l_2}(k_2,\eta_0)\Delta^{T,(1)}_{l_3}(k_3,\eta_0) 
P(k_2) P(k_3) \\ \nonumber   
&& \left(
\begin{array}{ccc}
l_1 & l_2 & L \\
0 & 0 & 0
\end{array}
\right)
\left(
\begin{array}{ccc}
L & l' &\tilde l \\
0 & 0 & 0
\end{array}
\right)
\left(
\begin{array}{ccc}
\tilde l & l' &  l_3 \\
0 & 0 & 0
\end{array}
\right) \times \\    && \nonumber
\sum_{m',\tilde m,M}
\left(
\begin{array}{ccc}
l_1 & l_2 & L \\
-m_1 & -m_2 & -M
\end{array}
\right)
\left(
\begin{array}{ccc}
L & l' &\tilde l \\
M & m' & \tilde m
\end{array}
\right)
\left(
\begin{array}{ccc}
\tilde l & l' &  l_3 \\
-\tilde m & -m' & -m_3
\end{array}
\right) \ +\ (2\leftrightarrow 3)\ .
\end{eqnarray}
Because of rotational symmetry, the CMB anisotropy bispectrum can be written as\footnote{Note that $\Delta_{lm}$ and $a_{lm}$ are nonlinearly related, but as usual we neglect all additional terms not enhanced by $\delta_e$.} 
\begin{equation}\label{blllrotinv}  
B^{m_1m_2m_3}_{l_1l_2l_3}=\langle a_{l_1m_1}a_{l_2m_2}a_{l_3m_3} \rangle=\frac{(\pi/4)^{3/2}}{\sqrt{\prod_i (2l_i+1)}}\langle\Delta_{l_1m_1}\Delta_{l_2m_2}\Delta_{l_3m_3} \rangle= \left(
\begin{array}{ccc}
l_1 & l_2 & l_3 \\
m_1 & m_2 & m_3
\end{array}
\right) \times B_{l_1,l_2,l_3}\ ,
\end{equation}
where 
\be\label{eq:B_total}
B_{l_1,l_2,l_3}= B_{l_1,l_2,l_3}^{a}+B_{l_1,l_2,l_3}^{b}\ ,
\ee 
is the rotationally invariant bispectrum, which is split into two pieces coming from the two pieces of the source function. 
Here we have used that the multipoles of the CMB anisotropies $a_{lm}$ are given by:
\be\label{alm}
a_{lm}=\frac{1}{4}\sqrt{\frac{4\pi}{2l+1}}\int\frac{d^3 k}{(2\pi)^3}\Delta_{lm}(\vec{k},\eta_0)\ .
\ee
Using the formula for the sum over the product of two 3$j$ symbols given in \cite{functionsWolfram} and combining (\ref{3pt}) and (\ref{3pttemp}), we obtain that the piece sourced by $S^i$ is given by
\be
\label{eq: B from _b}
&& B^{i,m_1m_2m_3}_{l_1l_2l_3}=\mathcal{G}^{l_1l_2l_3}_{m_1m_2m_3}i^{l_1+l_2+l_3}\frac{1}{2\pi^{5/2}}\frac{(-1)^{l_2+l_3}}{64\pi\sqrt{(2l_2+1)(2l_3+1)}}\int_0^{\eta_0} d\eta
g(\eta) \\   \nonumber &&
\times\left(\sum_{l',\tilde l} \right.
\sqrt{(2 \tilde l+1)}(2 l'+1)
 i^{l_3-l'-\tilde l}
\left(
\begin{array}{ccc}
l_3 & l' & \tilde l \\
0 & 0 & 0
\end{array}
\right)^2 \\ \nonumber    &&\left.  \times \left[\int
dk_2 k_2^2  j_{l_2}(k_2 (\eta_0-\tilde\eta)) \tilde
S^{i,1}(k_2,\eta)
\Delta_{l_2}^{T,(1)}(k_2,\eta_0)P(k_2) \right] \right.\\
   &&\nonumber \left.\times \left[\int dk_3 k_3^2
 j_{l'}(k_3 (\eta_0-\eta)) \tilde
S^{i,2}_{\tilde l}(k_3,\eta)
\Delta_{l_3}^{T,(1)}(k_3,\eta_0)P(k_3) \right]+ \ \ {\rm 5 \
symmetric\ terms \ }\right) \ .
\ee
which is a confirmation of the rotational symmetry of our formulas.

\section{Summary of the analytical results\label{sec:summary of analytical}}

Let us summarize the results of our analytical treatment. It is easy to realize that our second order expression for $\Delta^{(2)}$ in eq.~(\ref{deltalm2main}) exactly reproduces what we expected from eq.~(\ref{guess}) in the Introduction, which implies that also the form of the bispectrum is  very similar to the one given in sec.~\ref{sec:estimates}. In fact, if we start from part $(b)$ of the source, we can see that $\tilde S^{b,2}$ is nothing but the first order source, while $\tilde S^{b,1}$ is just equal to $\delta_g$:  part $(b)$ of the source reproduces exactly what we expected from the perturbation to the visibility function. Concerning the perturbation to the damping scale, this is given by part $(a)$ of the source. Also in this case,  $\tilde S^{a,2}/k^2$ turns out to reproduce nothing but the first order source (where we have not kept control of the metric perturbations, that therefore have disappeared). $\tilde S^{a,1}$ is just equal to $2\delta_{k_D}/k_D^2$. Putting all these pieces together, and obviously neglecting the complications coming from rotation invariance, we see that expression~(\ref{deltalm2main}) for $\Delta^{(2)}_{lm}$ reproduces what we had anticipated in eq.~(\ref{guess}).


The final expression we obtained for the bispectrum in  (\ref{eq:B_total}) is of course quite more complicated. A useful simplification is obtained if we express $\Delta_l^T$ given in (\ref{firstT}) using the standard photon temperature transfer function, $\Theta_l$ (given in e.g.~\cite{Dodelson} and which is returned from CMBFAST for example \cite{cmbfast}). Combining
\be\label{transferfunction}
\Delta_l^{T,(1)}(k,\eta_0)=(-1)^l 4\sqrt{4\pi(2l+1)}\Theta_l\ ,
\ee
where $T$ stands for transfer function, with (\ref{eq: legendre_to_spherical}), we see that $\Delta_l^{(1)}=(-i)^l 4 \Theta_l$. Using these transformations we can check that the multipoles of the CMB anisotropies defined in (\ref{alm}) reduce to the standard form  (up to an irrelevant phase):
\be\label{almdef}
a_{lm}=(-i)^{-l}\int\frac{d^3k}{(2\pi)^3}\int d\vec{n}Y_{lm}^*(\vec{n})\Theta(\vec{k},\vec{n},\eta_0)\ .
\ee

Further, since the source contains multipoles only up to the quadrupole, we have the following inequalities:  $0\le l_1''\le 2\ $,
$l_1-2\le l_1'\le l_1+2\ $, $l_3-2\le \tilde l_3\le l_3+2\ $. Combining these with the triangle inequalities for $l_1,l_2,l_3$, we can explicitely do the sums over $l_1',l_1'',\tilde l_3$.

Using (\ref{transferfunction}) we find~\footnote{One may wonder why our expression for the bispectrum contains two $k$-integrals, while the bispectrum in the local model contains 3 $k$-integrals. This is due to the fact that the perturbation $\delta_e$ affects the CMB only locally in real space: this is true for the effect on the visibility function along the line of sight, and approximately in the squeezed limit for the effect due to the damping. Technically, this boils down to the fact that the second order source factorizes as in eq.~(\ref{eq:source_redefinition}), which allows to eliminate one of the $k$ integrals using the two delta functions coming from (\ref{deltaf}).
}
\begin{eqnarray}\label{eq:B_total_simple}
&&\langle a_{l_1m_1}a_{l_2m_2}a_{l_3m_3} \rangle=\\ \nonumber
&&\ \
\  \ \ \ \ \ \
\mathcal{G}_{l_1l_2l_3}^{m_1m_2m_3}i^{l_1+l_2+l_3}\int_0^{\eta_0} dx
g(\eta_0-x) \left[\alpha^{a}_{l_2}(x)\beta^{a}_{l_3}(x)+(a\to
b)\right]+ {\rm 5
\ symm.}\ ,\\
&&\alpha^{i}_{l_2}\equiv\frac{2}{\pi}\int dk_2 k_2^2
\Theta_{l_2}(k_2,\eta_0)P(k_2) j_{l_2}(k_2 x_{\tilde\eta}) \tilde
S^{i,1}(k_2,\eta_0-x)\ ,\\ 
&&\beta^{i}_{l_3}\equiv\frac{1}{8\sqrt{\pi}}\frac{2}{\pi}\int dk_3
k_3^2 \Theta_{l_3}(k_3,\eta_0)P(k_3)\times\\ 
&&\left( j_{l_3}(k_3 x) \tilde
S^{i,2}_0(k_3,\eta_0-x)
+\frac{\sqrt{3}}{2l_3+1}
\left[l_3j_{l_3-1}(k_3 x)-(l_3+1)j_{l_3+1}(k_3 x)\right] \tilde
S^{i,2}_1(k_3,\eta_0-x)+\right.\nonumber\\
&&\left.+\frac{\sqrt{5}}{2}\left[3j''_{l_3}(k_3
x)+j_{l_3}(k_3x)\right]\tilde S^{i,2}_2(k_3,\eta_0-x)
\right)\ . \nonumber
\end{eqnarray}
where $x_{\tilde\eta}=x$ for all terms in the source, except the term containing the integral over $\delta_e$ in $\eta'$, in which case $x_{\tilde\eta}=\eta_0-\eta'$, and $j_{l_2}$ must be brought inside the integral over $\eta'$ in $\tilde
S^{b,1}$.
This is our final expression for the bispectrum generated during recombination, which we will evaluate in the next section. Notice that in the limit in which the time dependence of $\delta_e$ is slow compared to the width of the visibility function, $\alpha_l^{a}$ and  $\alpha_l^{b}$ can be evaluated at $x=\eta_r$ and approximately brought out of the $x$ integral. What remains is proportional to the variation of $C_{l_3}$ with respect to $\log X$, where $X$ are the quantities perturbed by $\delta_e$, reproducing the approximate expression for the bispectrum  found in (\ref{eq:approx_bispectrum}).

\section{Results and Discussion\label{sec:resultsand discussion}}

\subsection{Signal-to-noise of the bispectrum}

We evaluate numerically eq.~(\ref{eq:B_total_simple}), working in synchronous gauge, using the following cosmological parameters:
$$(\Omega_b,\,\Omega_\Lambda,\,h,\,T_{\rm cmb}\,,Y_p,\,n_s,\tau)=(0.0441,\,0.742,\,0.719,\,2.725,\,0.24,\,0.963,\,0.087)\, ,$$
and we are now ready to compute the signal-to-noise of the bispectrum induced by perturbations to the recombination history.  We concentrate on a full sky cosmic variance limited experiment between $l_{\rm min}$ and $l_{\rm max}$, without accounting for lensing~\footnote{This is a good approximation up to $l$ of order 3000}. The Fisher matrix between two bispectra $i$ and $j$ is given by (see for example \cite{komatsu}):
\be\label{Fish}
F_{ij}\equiv \sum\limits_{l_{\rm min}\leq l_1 \leq l_2 \leq l_3 \leq l_{\rm max}}\frac{B^{(i)}_{l_1l_2l_3}B^{(j)}_{l_1l_2l_3}}{C_{l_1}C_{l_2}C_{l_3}\Delta_{l_1l_2l_3}}\ ,
\ee
where  $\Delta_{l_1l_2l_3}=1,2,6$ for triangles with no, two or three equal sides; and $C_l$ is the CMB angular power spectrum. We decide to  concentrate only on the bispectra from recombination and from the local model (with $f_{\rm NL}^{\rm loc.}=1$). The local model is a good representative for the non-gaussianities one can expect from inflation (in this case multi-field inflation \cite{Creminelli:2004yq, Cheung:2007sv}
), and therefore it offers a qualitative measure of the degradation between the primordial and the cosmological signals~\footnote{Since we will find that the degradation on the local model due to recombination is relatively small, we can avoid to compute explicitly the degradation induced on the equilateral model, which can come from single field inflation.}. 

The signal-to-noise ratio $S/N$ for the recombination bispectrum is given by
\be\label{SoN}
\left(S/N\right)^2= F_{rec,rec}\ .
\ee
This assumes a negligible degradation from the local model~\footnote{If the correlation with the local model is included in calculating the $S/N$, the result for $l_{\rm min}=100$ and $l_{max}=3000$ is changed only by 2\%.}.

We plot the total $S/N(<l_{\rm max})$ as a function of $l_{\rm max}$ in Fig.~\ref{fig:sn}. The minimum $l$ we include in the calculation is $l_{\rm min}=100$. This means that, consistently with our approximation, we sum over modes inside the horizon at recombination, which corresponds to $l\approx70$. We obtain $S/N\approx 0.4$ for $l_{max}=3000$. This means that such a signal should not be detectable by a satellite like Planck from the analysis of the temperature signal alone. However, the information contained in the polarization signal is expected to be comparable to the one in the temperature  \cite{2004PhRvD..70h3005B}, and therefore we expect that if our calculation were extended to  include polarization, the resulting signal may be Planck detectable. The $f_{\rm NL}^{\rm loc.}$ which in the local model generates the same signal-to-noise for the given $l_{\rm min}$ and $l_{\rm max}$ is usually referred to as $f_{\rm NL}^{\rm eff.}$ and thus our result corresponds to 
\be
f_{\rm NL}^{\rm eff.}\approx -3.5\pm\mathcal{O}(1)\ ,
\ee
where the correction $\mathcal{O}(1)$ comes from what we naively expect from the terms not enhanced by $\delta_e$ that we dropped. The effective $f_{\rm NL}^{\rm loc.}$ is a convenient way to parametrize the total signal-to-noise but one should keep in mind that it carries no information about the shape of the bispectrum. The sign of $f_{\rm NL}^{\rm eff}$ is chosen according to the convention discussed below.  We also compute the correlation coefficient between the bispectra induced by recombination and by the local model, $r_{rec,loc}\equiv F^{-1}_{rec,loc}/\sqrt{F^{-1}_{rec,rec}F^{-1}_{loc,loc}}$ as a function of $l_{\rm max}$. This is shown in Fig.~\ref{fig:sn}, where one can see that for low $l_{\rm max}$ it is equal to -1, and it decreases rapidly as we include higher $l$'s. The Fisher matrix with the local model, for $l_{\rm max}=3000$, is given by:
\bea
F=\left(
\begin{array}{cc}
0.19 & 0.011 \\
0.011 & 0.016
\end{array}
\right)\ ,
\eea
where index 1 corresponds to the non-Gaussianities from recombination, while index 2 corresponds to  the local model.  From this we can deduce that the bias of the non-Gaussianities from perturbing recombination into the estimate of $f_{\rm NL}^{\rm loc.}$ is about 0.7 at this resolution. 



In Fig.~\ref{fig:sn} we also plot the contribution to $S/N$ from each $l_{max}$--, and each $l_{min}$--bin by plotting~$d(S/N)^2/dl_{\rm min}$ and $d(S/N)^2/dl_{\rm max}$. It is easy to see that the biggest contribution to the total $S/N$ comes from squeezed triangles with 
\be\label{config}
l_2 \approx l_3\gtrsim2000,\ l_1\sim 200\ ,
\ee
i.e. with the small $l$ around the first acoustic peak, and the two high $l$s in the diffusion damping tail. This result is consistent with our expectations stated at the beginning of sec.~\ref{sec:estimates} based on considering the timescales at recombination. The fact that the signal is not dominated by triangles with the smallest of the $l$'s we consider (it decays for $l\lesssim 200$) is a confirmation of the viability of our approximation of neglecting metric perturbations. This result matches what was found explicitly in \cite{2004PhRvD..70h3532C} for triangles  with the smallest side out of the horizon.

\begin{figure}[h!]
\begin{center}
\includegraphics[width=15cm]{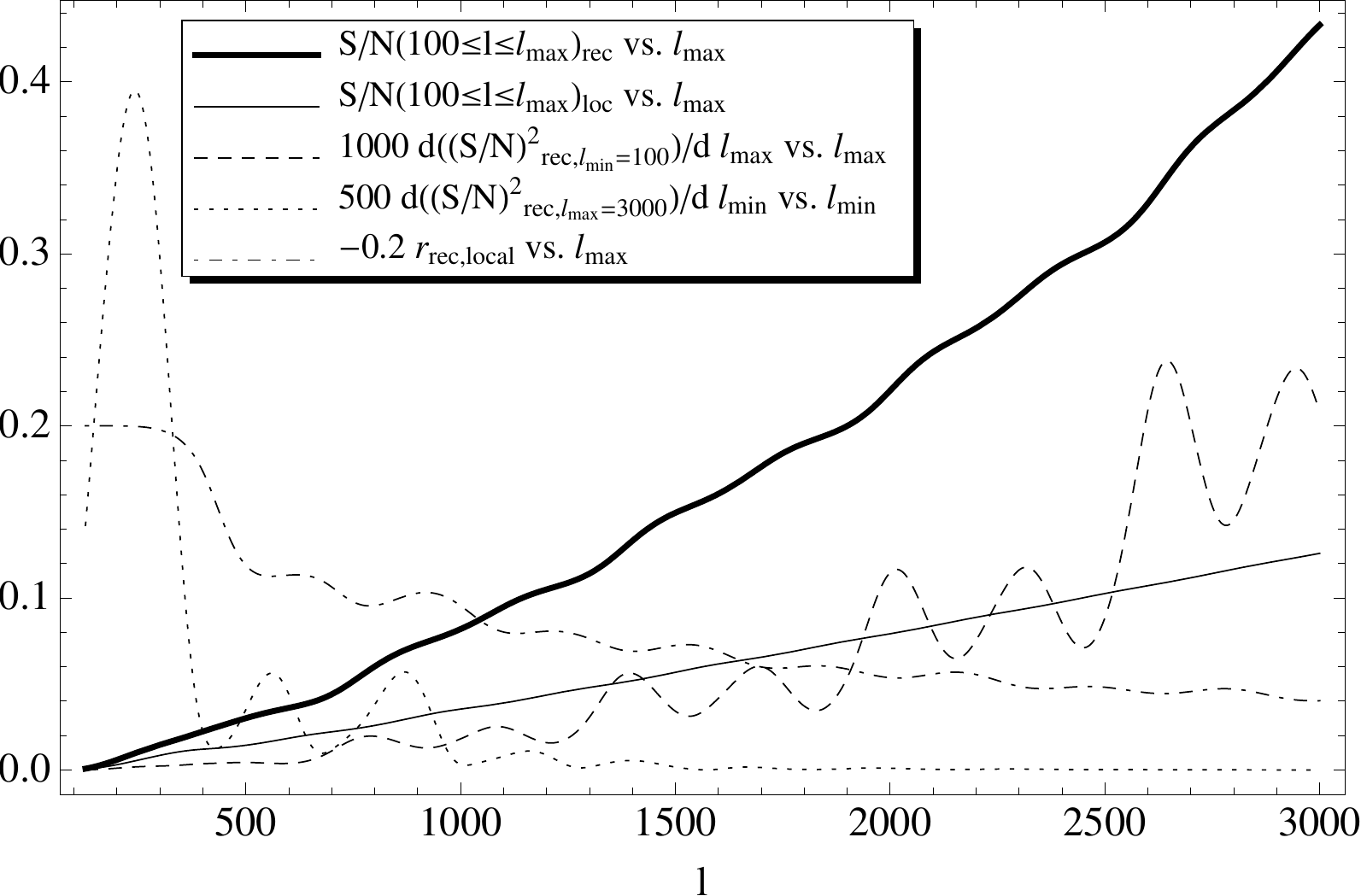}
\end{center}
\caption{\small The plot shows the signal-to-noise for the recombination and local model bispectra; $d(S/N)^2/d l_{max}$ and $d(S/N)^2/d l_{min}$ (for triangles with $l_{min}=100$ and $l_{max}=3000$, respectively) of the bispectrum generated around recombination; and the correlation coefficient between the local model and recombination vs $l_{\rm max}$ with $l_{\rm min}=100$. The computation was done for an ideal (cosmic variance limited) experiment without accounting for lensing. Clearly triangles with two high $l$s, and a low-$l$ around 200 dominate the recombination signal-to-noise. The local model and recombination bispectra start decorrelating only after high-$l $ modes are included.
} \label{fig:sn}
\end{figure}

\subsection{The shape of the non-Gaussianities}

Let us discuss the shape of the non-gaussian signal produced by fluctuations in $\delta_e$.
The signal-to-noise per triangle, which is given by  $|B_{l_1l_2l_3}|/\sqrt{C_{l_1}C_{l_2}C_{l_3}}$, can be read off from eq.~(\ref{SoN}). It includes the additional weighting from the 3-$j$ symbols which takes into account the phase space distribution of the triangles. The physical information in the bispectrum is contained in the $S/N$ per triangle divided by the phase space weight of that triangle shape, which gives $b_{l_1l_2l_3}/\sqrt{C_{l_1}C_{l_2}C_{l_3}}$ (we prefer to preserve the information encoded in the sign of $b$). $b_{l_1l_2l_3}$ is the reduced bispectrum and is defined as $B_{l_1l_2l_3}^{m_1m_2m_3}\equiv\mathcal{G}_{l_1l_2l_3}^{m_1m_2m_3} i^{l_1+l_2+l_3} b_{l_1l_2l_3}$. It can be read off from (\ref{eq:B_total_simple})~\footnote{ The bispectrum is defined up to a phase. Our definition of the bispectrum contains an extra factor of $i^{l_1+l_2+l_3}$ (which is real since $l_1+l_2+l_3$ is even) compared to \cite{komatsu} due to the extra factor of $i^l$ in our definition of $a_{lm}$ (\ref{almdef}).} . The reduced bispectrum reduces to the bispectrum in the flat-sky approximation \cite{komatsu}, and thus it contains the physical information of the three-point function. 
 
 Let us comment briefly on the sign of the bispectrum. With our definition, the signs of $B_{l_1l_2l_3}$ and $b_{l_1l_2l_3}$ are the same. These are chosen such that they coincide with the one of the reduced bispectrum in \cite{komatsu} in the ``local'' model, in which a positive $f_{\rm NL}^{\rm loc.}$ parameter corresponds to a negative skewness of the one-point distribution function of the temperature anisotropies, and to a negative reduced bispectrum in the Sachs-Wolfe regime (see eq.~4.24 of~\cite{komatsu}). With our definition, a generally positive bispectrum (negative $f_{\rm NL}^{\rm eff.}$) corresponds to high-$l$ modes having enhanced amplitude in hot long-wavelength patches of the sky.

Let us now analyze the shape of the signal. In Fig.~\ref{fig:l1700a} we plot $l_3^{3/2}\times b_{l_1l_2l_3}/(C_{l_1}C_{l_2}C_{l_3})^{1/2}$, for $l_1\leq l_2\leq l_3$ (to avoid redundancy) subject to the triangle inequalities for a typical high-$S/N$ $l_3$. In the plot we choose $l_3=1700$. 
For comparison, we plot the analogous quantity in the case of the local model. The factor of $l_3^{3/2}$ is introduced so that after normalization of the $l_1$ and $l_2$ axes by dividing by $l_3$, the integral over the square of the plotted function, multiplied by the phase space weight (see eq.~(4.54) of \cite{komatsu}), is equal to the $(S/N)^2$  per logarithmic $l_{\rm max}=l_3$ bin. The most important feature to notice is that, as anticipated, the signal is peaked on squeezed triangles, but the shape is different from the one of the local model.

\begin{figure*}[h!]
\begin{minipage}{155mm}
(a) \hskip 68mm (b)
\begin{center}
\epsfig{file=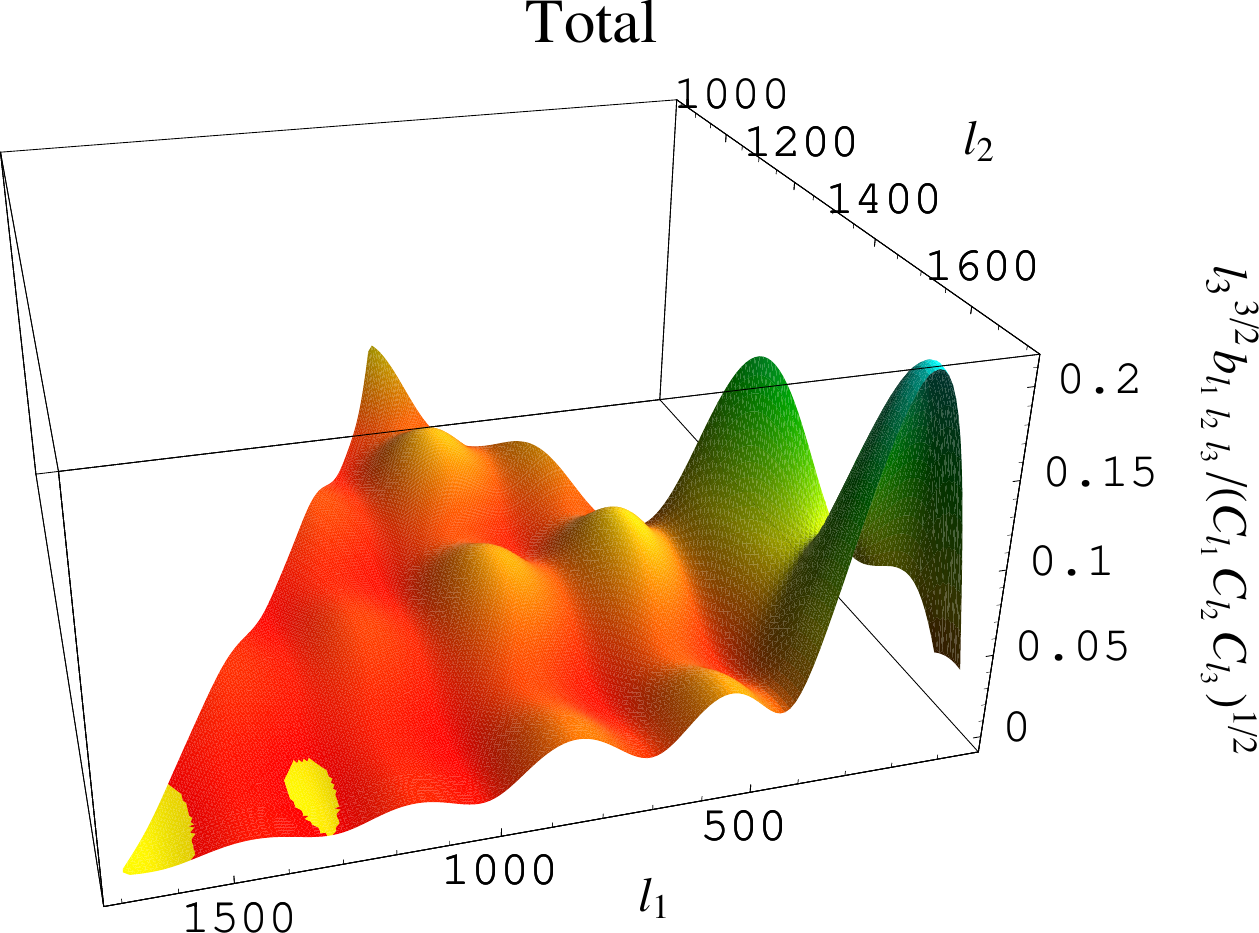, width=7.2cm,bbllx=0,bblly=0,bburx=360,bbury=270}
\epsfig{file=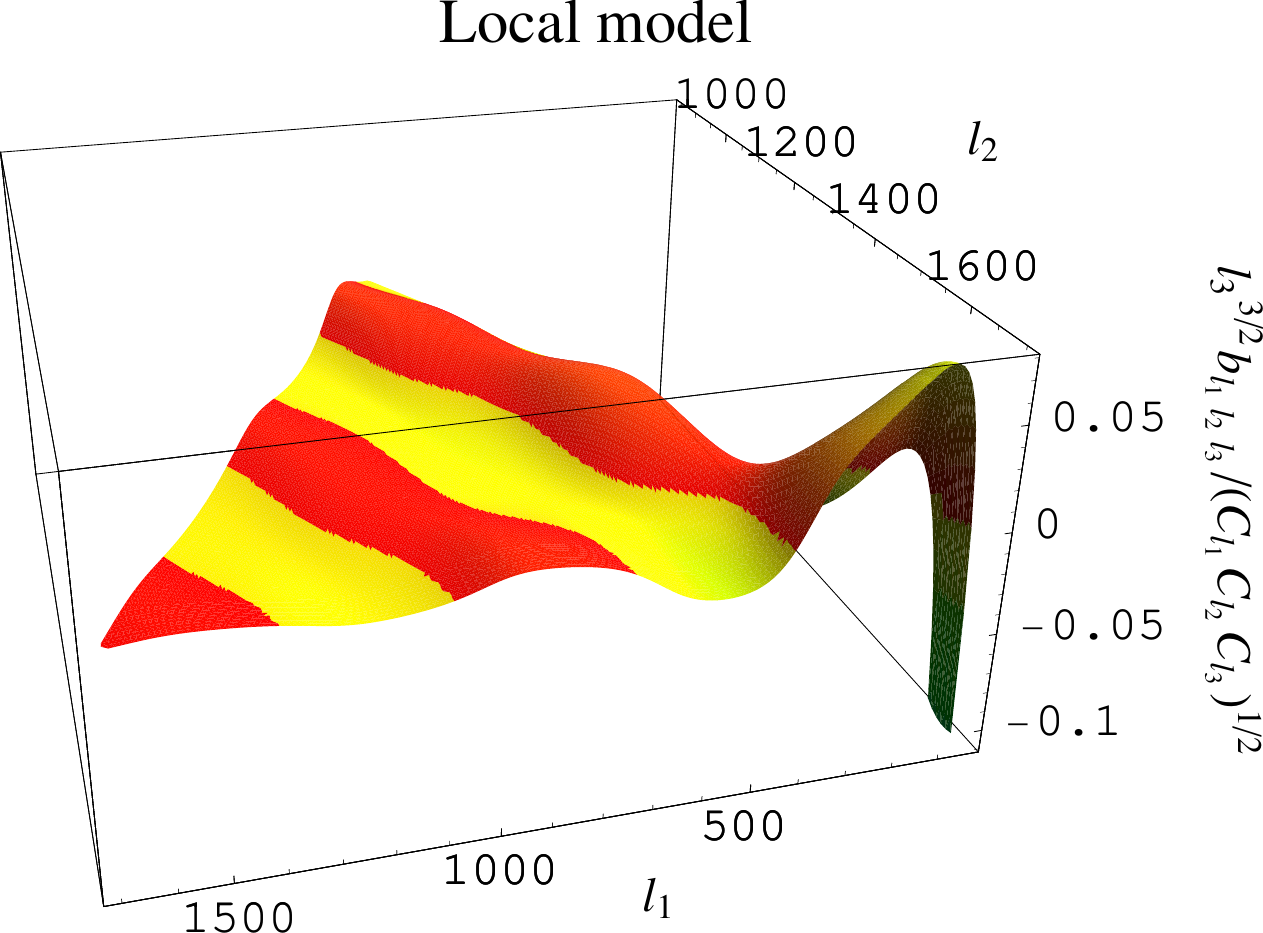, width=7.2cm,bbllx=-20,bblly=0,bburx=340,bbury=270}
\end{center}
\caption{\small 
In panel (a) we show $b_{l_1l_2l_3}^{rec.}/(C_{l_1}C_{l_2}C_{l_3})^{1/2}$ for an high-$S/N$ $l_3$ (we choose $l_3=1700$). For comparison, in panel (b) we plot the analogous quantity from the local model with $f_{\rm NL}^{\rm loc.}=1$. The peak in the lower-right corner of panel (a) ($l_1\approx 200$, $l_2\approx l_3=1700$) corresponds to the high-$S/N$ squeezed triangles given by (\ref{config}). These triangles are such that they sample both the Silk damped tail, and the first acoustic peak. As for the local model, the signal is dominated by squeezed triangles, though the shape is clearly different. The acoustic oscillations can be clearly seen. The color scheme of the plots is chosen to highlight $b_{l_1l_2l_3}=0$. }\label{fig:l1700a}
\end{minipage}
\end{figure*} 

For purposes of comparison with the literature (see for example \cite{komatsu}), we plot in Fig.~\ref{fig:blll} the reduced bispectrum as a function of the smallest $l$, $l_1$, for isosceles configurations with the higher $l$'s, $l_2=l_3$, ranging from $\sim2200$ to $\sim 2300$. As $l_1$ changes, we clearly see the acoustic oscillations in the bispectrum. The change in value as $l_2$ varies is also due to the acoustic oscillation.
 

\begin{figure}[h!]
\begin{center}
\includegraphics[width=15cm]{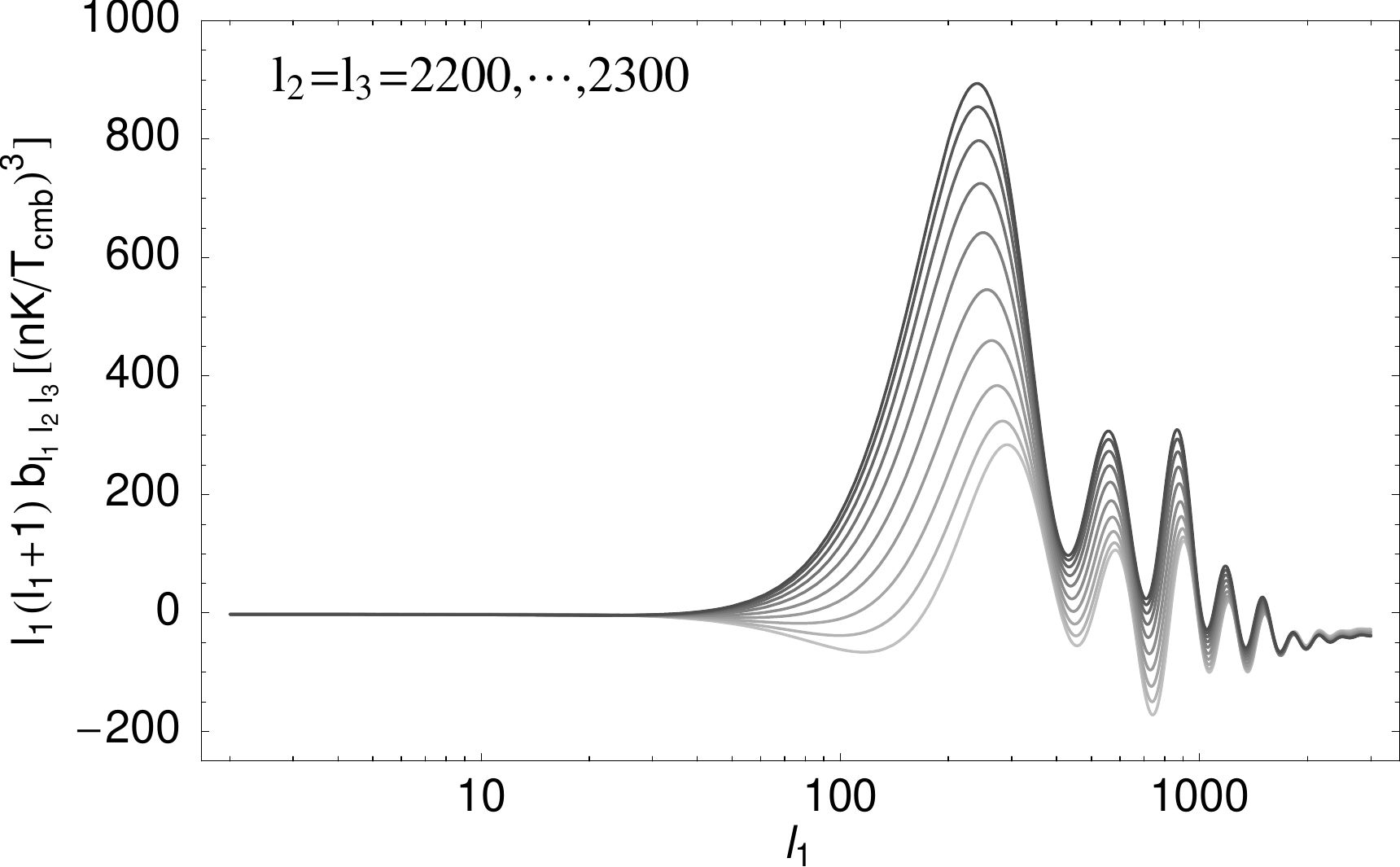}
\end{center}
\caption{\small The reduced bispectrum for triangles with $l_2=l_3$. The different curves correspond to equally spaced values of $l_2=l_3$. The bottommost (light-gray) curve is for $l_2=l_3=2200$, while the topmost (dark-gray) curve is for $l_2=l_3=2300$. The similarity between the bispectrum for these triangles and the temperature anisotropy power spectrum is apparent.} \label{fig:blll}
\end{figure}

In order to get some intuition on how much the signal is scale invariant, in Fig.~\ref{fig:shapes} we plot the $S/N$ per triangle for isosceles triangles ($l_2=l_3$). Each curve represents a triangle with a certain level of squeezing,  characterized by the ratio $l_1/l_2$, with $l_1\leq l_2$, as we vary the overall size of the triangle characterized by $l_2$.  The topmost curve corresponds to the most squeezed triangles, which confirms that most of the signal is in that kind of triangles. However, we see that as the size of the triangle varies, the amplitude of the signal fluctuates a lot. This is a violation of scale invariance. In fact, the thick curve represents the analogous quantity for a local shape with $f_{\rm NL}^{\rm loc.}=1$ for the maximum squeezing, whose signal is generated by primordial scale invariant perturbations.
 We find that the equilateral and obtuse triangles have a completely negligible contribution. 
 Only squeezed triangles with ratio $l_1/l_2\lesssim 0.1$ have a signal larger than the ones corresponding to $f_{\rm NL}^{\rm loc.}=1$.
 
 \begin{figure}[h!]
\begin{center}
\includegraphics[width=15cm]{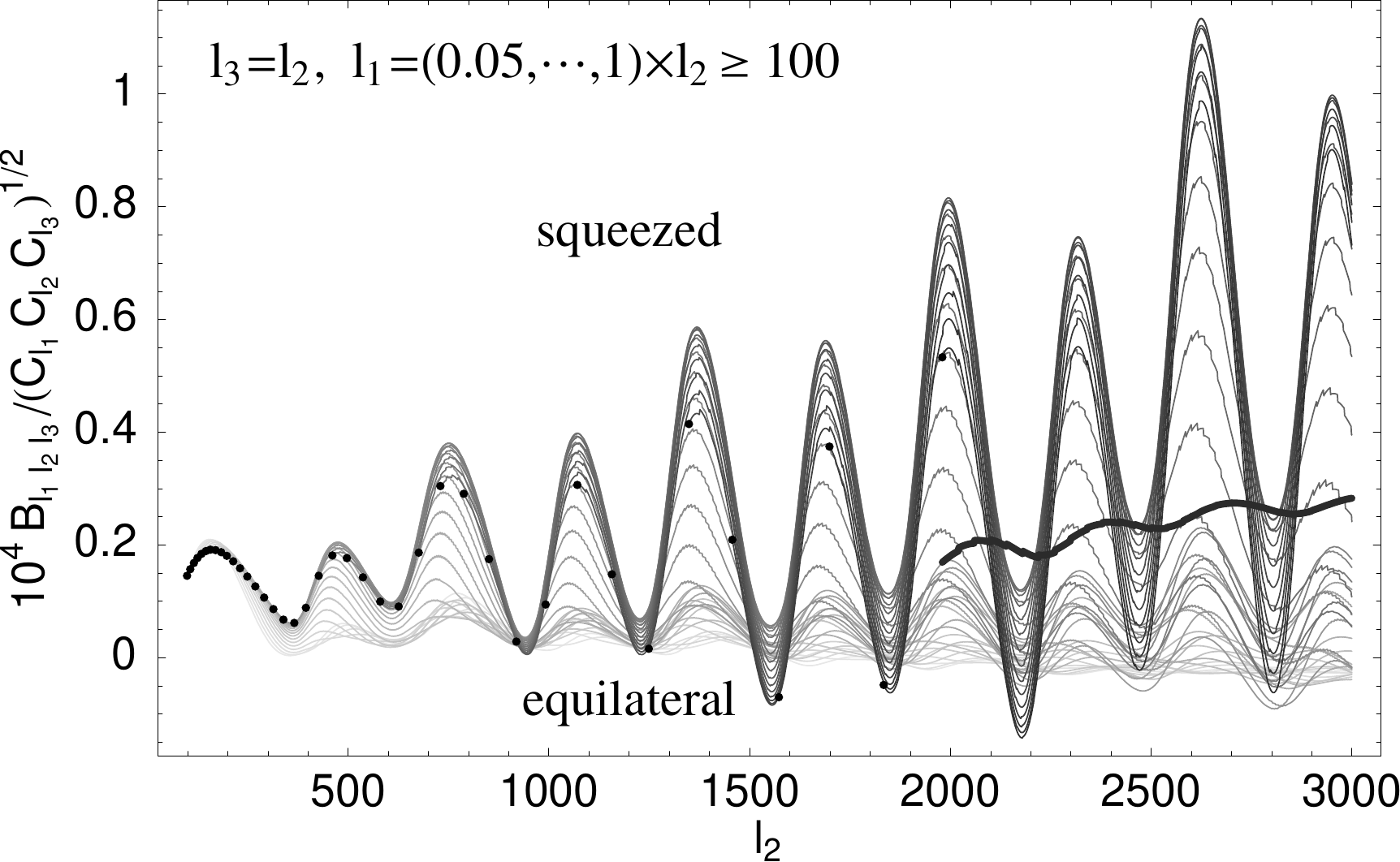}
\end{center}
\caption{\small The $S/N$ per triangle for triangles of various shapes with $l_2=l_3$. The different curves correspond to different ratio of $l_1/l_2$ which varies between 0.05 and 1 in equal logarithmic steps. Each curve terminates on the left with a dot, where $l_1=100$, which approximately is the minimum $l$ below which we can not trust our solution anymore. On the rightmost side of the plot, there are 40 curves, and this number is reduced to the left. The bottommost (light-gray) curve corresponds to equilateral triangles, while the topmost (dark-gray) curve is for highly squeezed triangles with $l_1/l_2=0.05$. The thick curve represents the analogous quantity for the local model with $f_{\rm NL}^{\rm loc.}=1$ and $l_1/l_2=0.05$. From the plot we can see that only extremely squeezed triangles with ratio $l_1/l_2\lesssim 0.1$ have enhanced $S/N$. Obtuse triangles have negligible contribution, though we do not plot them explicitly. } \label{fig:shapes}
\end{figure}

\subsection{Physics of the bispectrum}

In sec.~\ref{sec:estimates}, we provided simple estimates for the origin and the size of the effects of the fluctuations of $\delta_e$ on the bispectrum. After the full calculation, we are now able to check those results.
From Fig.~\ref{fig:sn}, we saw that most of the signal comes from squeezed triangles with $l_1\ll l_2\simeq l_3$. In that limit, the leading contributions to $b_{l_1l_2l_3}$ come from the terms $\alpha^{a}_{l_1}(\beta^{a}_{l_{2}}+\beta^{a}_{l_{3}})$ and $\alpha^{b}_{l_1}(\beta^{b}_{l_{2}}+\beta^{b}_{l_{3}})$  of eq.~(\ref{eq:B_total_simple}). These terms respectively correspond to the following combinations: low-$k$ $\delta_{k_D}$ and high-$k$ first order source $S^{(1)}$; and low-$k$ $\delta_g$ and high-$k$ $S^{(1)}$. The other interesting term, high-$k$ $\delta_g$ and low-$k$ $S^{(1)}$, gives a negligible contribution as we can expect from our estimates. This is exactly the same structure we found in sec.~\ref{sec:estimates}. In particular, the fact that the highest signal from perturbations to the diffusion scale due to $\delta n_e$, which we call $\delta_{k_D}$, comes from squeezed triangles with a slowly varying $\delta_e$, is a consistency check of our approximate treatment of the diffusion damping at second order in sec.~\ref{sec: part a}. 
The contribution from the diffusion damping perturbation to the total $S/N(<l_{max})$ from triangles for which that approximation breaks down is $\sim10\%$, which is equivalent to a contribution of $\Delta f_{\rm NL}^{\rm eff.}\sim 1/2$ to the bispectrum. These order one corrections to $f_{\rm NL}^{\rm eff.}$ are of the order of the ones we expect from terms we neglect such as second order corrections from metric perturbations, and are therefore beyond our control.

\begin{figure}[h!]
\begin{center}
\includegraphics[width=15cm]{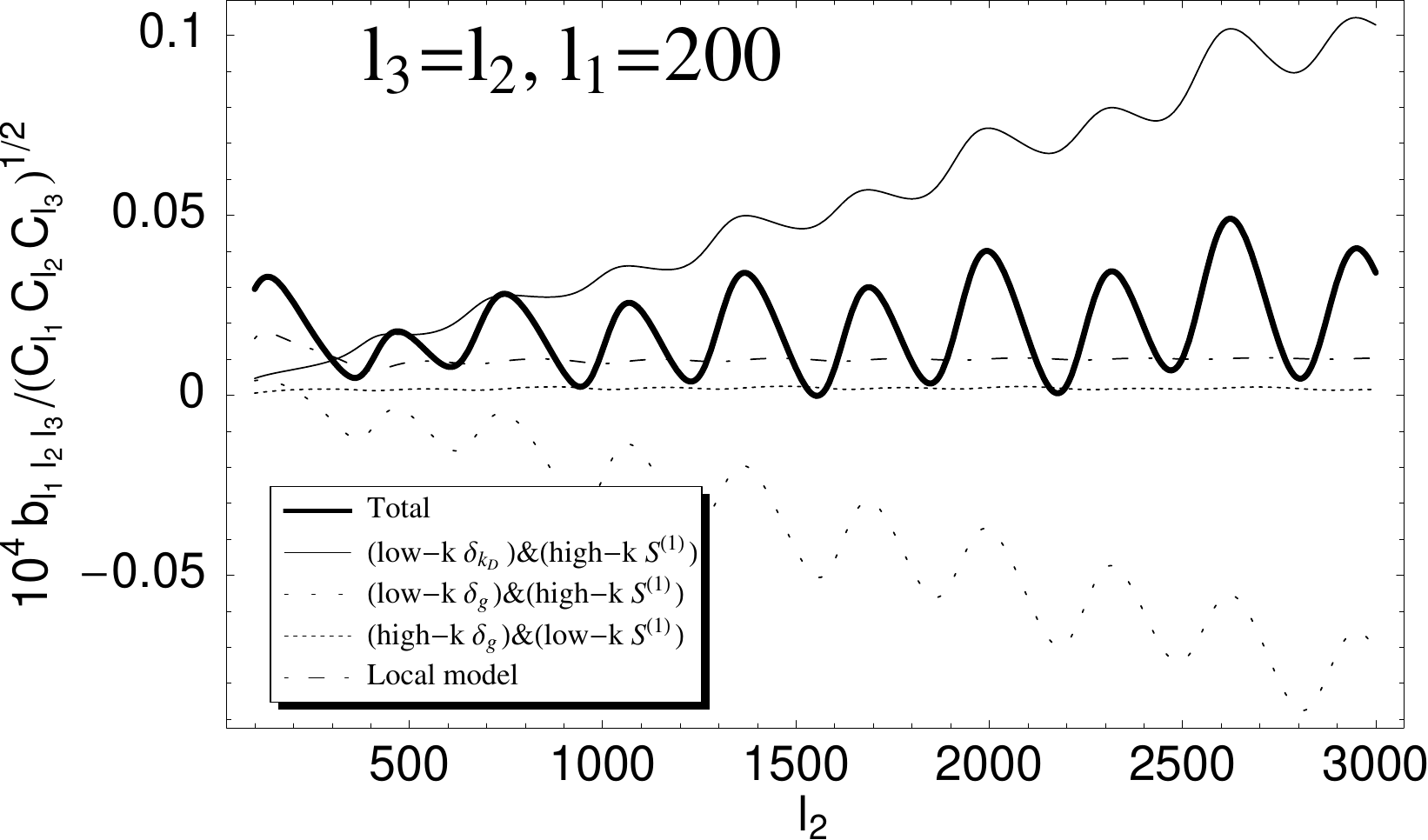}
\end{center}
\caption{\small The $S/N$ per triangle (keeping the sign of $b$) for typical high-$S/N$ squeezed triangles with $l_2=l_3$ and $l_1=200$. We also show separately the three important contributions to $S/N$. Notice the similarity between this plot and the plot obtained using the approximate treatment of the bispectrum in the squeezed triangle limit, Fig.~\ref{fig:btot}. We reproduce all qualitative features of that plot, which serves as a check of our code, and allows us to pinpoint the main physical effects generating the bispectrum (see the text). } \label{fig:sq}
\end{figure}

In Fig.~\ref{fig:sq} we plot the three contributions discussed above to the $S/N$ per triangle for  high-$S/N$ squeezed triangles.  The bispectrum generated from $\delta_{k_D}$ is generally positive and it dominates the one from $\delta_g$ which is generally negative. Thus, the total bispectrum from recombination is generally positive, i.e. the amplitude of the high-$l$ modes is enhanced in hot long-wavelength patches of the sky corresponding to the first acoustic peak scale. Notice that $\delta_g$ receives contributions from perturbations  to $k_D$ due to the shift in the position of the last scattering surface ($\delta_{k_g}$ in the notation of sec.~\ref{sec:estimates}), from perturbations to the phase of the modes at the last scattering surface, and from perturbations to  the probability for the CMB photons to originate 
from different times within the recombination era ($\delta {\rm Area}$ still in the notation of sec.~\ref{sec:estimates}). As we anticipated, the leading contributions from $\delta_{k_g}$ and $\delta_{k_D}$ cancel in the limit in which $\delta_e$ is much slower than the timescale of recombination, which explains the partial cancellation between the two terms seen in Fig.~\ref{fig:sq}. This can be seen also in Fig.~\ref{fig:l1700b}, where we plot the analogous of Fig.~\ref{fig:l1700a} for the same three contributions.

\begin{figure*}[h!]
\begin{minipage}{155mm}
(a) \hskip 68mm (b)
\begin{center}
\epsfig{file=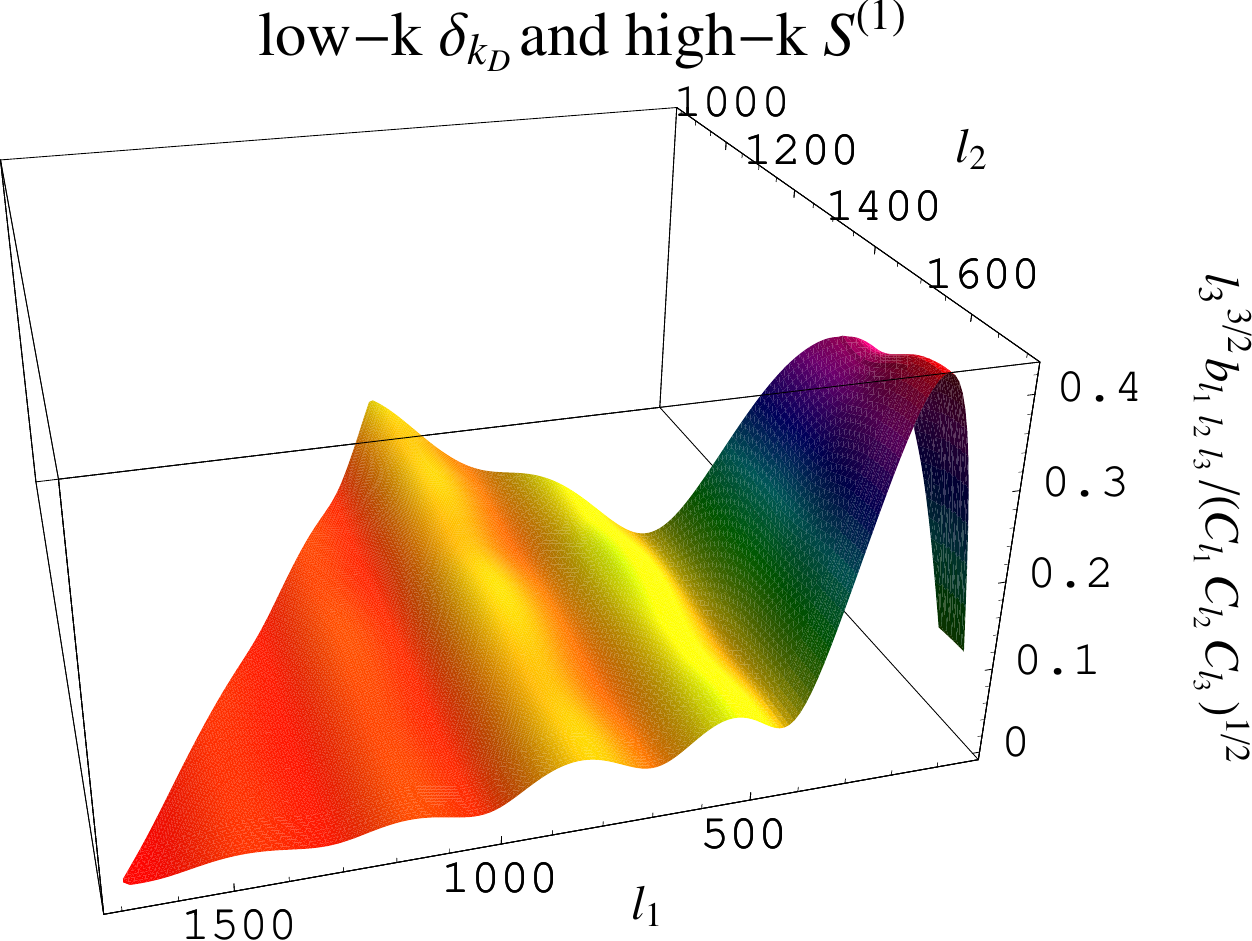, width=7.2cm,bbllx=-20,bblly=0,bburx=340,bbury=270}
\epsfig{file=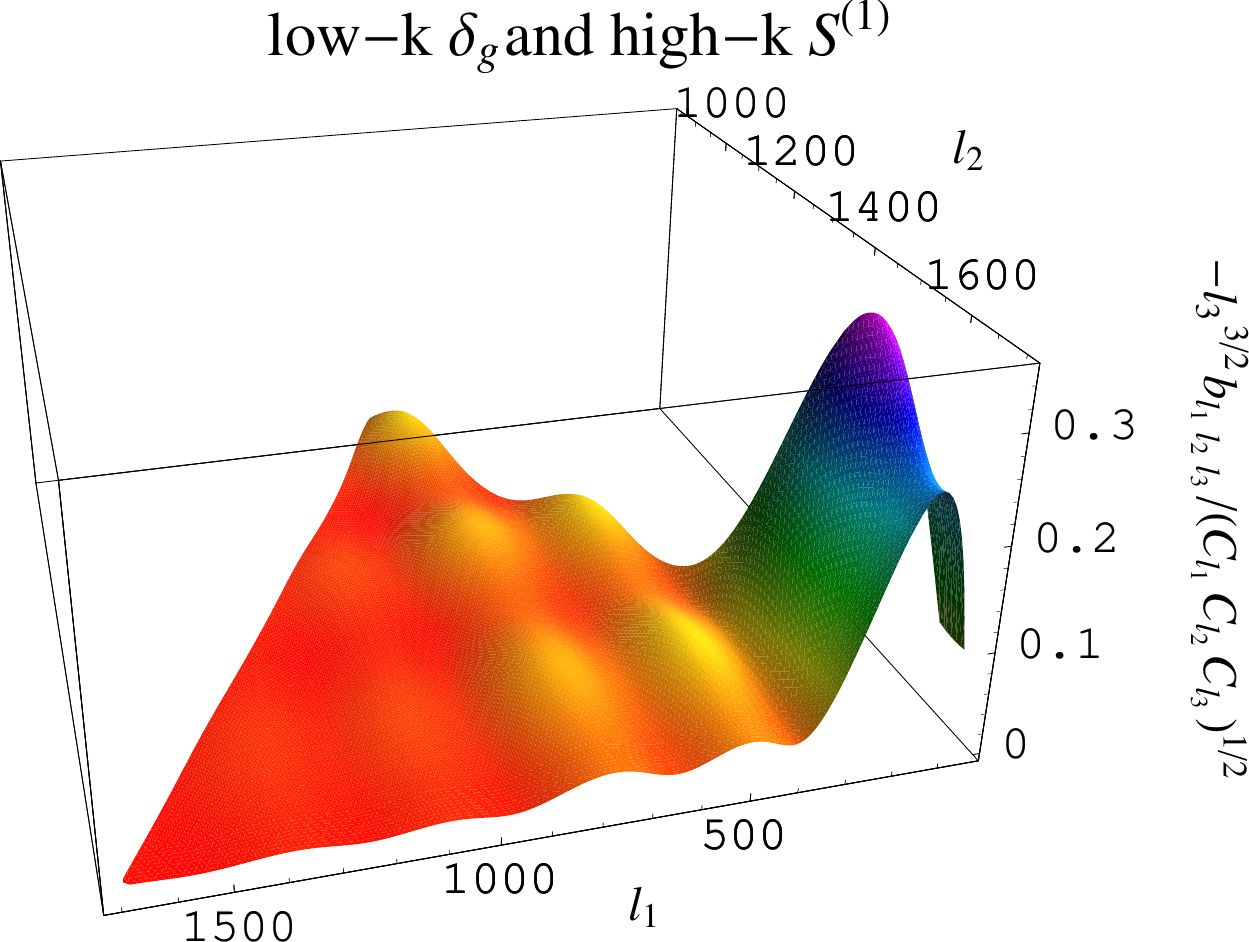, width=7.2cm,bbllx=-20,bblly=0,bburx=340,bbury=270}
\end{center}
 \hskip 40mm  (c)
\begin{center}
\epsfig{file=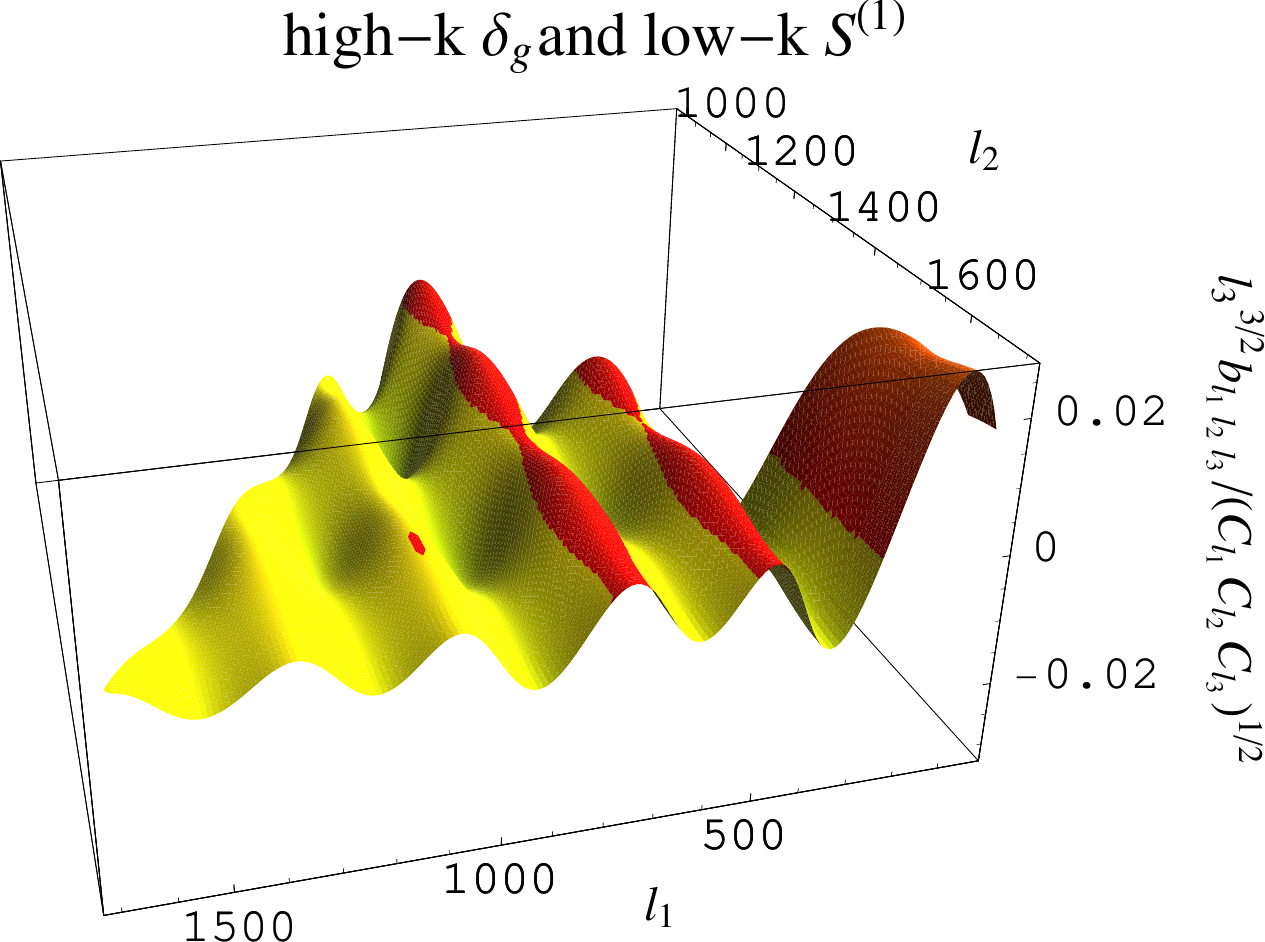, width=7.2cm,bbllx=0,bblly=0,bburx=360,bbury=270}
\end{center}
\caption{\small The same plot as in Fig.~\ref{fig:l1700a} for the three interesting contributions to the recombination bispectrum. a) Bispectrum generated by low-$k$ $\delta_{k_D}$; b) Bispectrum generated by low-$k$ $\delta_{k_g}$; c) Bispectrum generated by high-$k$ $\delta {\rm Area}$. Each of the two main contributions is peaked in the squeezed limit. Note that, as anticipated, the contribution to the bispectrum from the perturbations to the visibility function is opposite in sign to the contribution from diffusion damping (see vertical-axis label). }\label{fig:l1700b}
\end{minipage}
\end{figure*} 

If we compare Fig.~\ref{fig:btot} from our approximate treatment of sec.~\ref{sec:estimates} with the same one from the full calculation, Fig.~\ref{fig:sq}, we notice that the semianalytical treatment reproduces with good accuracy the general trends and the oscillation frequencies, phases and relative amplitudes of the various terms. The additional averaging coming from the integration in time of the visibility function in the full calculation suppresses the bispectrum generated by $\delta$-phase in the full calculation with respect to the semianalytical result, as expected. Very importantly, this tells us that we can trust both our code and the semianalytical analysis, and pinpoint the main physical effects generating the recombination bispectrum, as described in sec.~\ref{sec:estimates}. In particular, the low-$k$ $\delta_g$ has a larger oscillation amplitude than the contribution from the low-$k$ $\delta_{k_D}$ that is due to the perturbation to the phase of the first order source (see  Fig.~\ref{fig:bL}) entering in (\ref{blllsecond}). The general upward trends (in magnitude) of the low-$k$ $\delta_{k_D}$ and $\delta_g$ are due to perturbations to the diffusion damping generated by $\delta n_e$ and $\delta\eta_r$, respectively. By comparing the shape of the signal induced by a high-$k$ $\delta_e$ with what we obtained in Fig.~\ref{fig:bH}, we can confirm that the high-$k$ $\delta_g$ bispectrum is negligible. Overall, the sum of all these contributions gives a bispectrum which has an $f_{\rm NL}^{\rm eff.}\simeq -3.5$, which is enhanced from the first naive expectations of $f_{\rm NL}^{\rm eff.}\sim 1$.

\section{Summary}\label{sec:summary}

The calculation of the bispectrum induced on the CMB by the standard cosmological evolution has become necessary to fully exploit the signal measured by the next generation CMB experiments such as Planck.

In this paper we have computed in an approximate way the bispectrum due to perturbations in the recombination history. In a companion paper \cite{inprep1} we computed the first order perturbations to the free electron density $\delta_e$, which is one of the relevant quantities for studying the perturbed recombination history. This quantity does not enter in the first order calculation of the CMB temperature fluctuations which allows us to predict the power spectrum, but instead it enters at non-linear level, and therefore it can induce a bispectrum. What we found is that the perturbation to $n_e$ is about a factor of 5 larger than the other perturbations, which implies the possibility for these fluctuations to generate a bispectrum with $f_{\rm NL}^{\rm eff}\sim 5$, about a factor of 5 larger than naive expectations, and possibly detectable by next generation experiments.

We have therefore set up to make this computation. The full second order calculation of the CMB bispectrum is a very hard task, with many terms that source the second order matter, radiation, and metric perturbations. By concentrating on modes well inside the horizon at recombination, we expect to be able to neglect the second order metric perturbations. This means that our results will be wrong by an effective $f_{\rm NL}^{\rm eff.}\sim{\cal{O}}(1)$, which is the signal expected to be induced by metric perturbations~\footnote{Recent estimates in \cite{Pitrou:2008ak} and \cite{Bartolo:2008sg} gives an $f_{\rm NL}^{\rm equil.}\sim{\cal{O}}(10)$ for equilateral configurations, which corresponds to roughly   $f_{\rm NL}^{\rm eff.}\sim{\cal{O}}(3)$.}. Because of the large value of $\delta_e$, we have concentrated on those second order source terms that are proportional to $\delta_e$, and that therefore are expected to give an enhanced effect with respect to the rest.

What we find is that the induced bispectrum on the temperature signal corresponds to an $f_{\rm NL}^{\rm eff.}\simeq-3.5\pm{\cal{O}}(1)$. The signal is peaked on squeezed triangles with the lowest $l\sim200$ and the highest ones at about $l\gtrsim 2000$,  and it is very far from being scale invariant. Such an $f_{\rm NL}^{\rm eff.}$ corresponds to a signal-to-noise for an experiment like Planck of order one half. It is expected that the polarization signal will contain an amount of information comparable to the one in the temperature, and therefore such a signal may be detectable by such an experiment.

We find that the physical origin of the signal can be understood quite simply in terms of three physical effects each one giving an $f_{\rm NL}^{\rm eff.}$ of order a few. The first effect is due to the time shift $\delta \eta_r$ induced on the time of recombination by a low-$k$ $\delta_e$ mode. This induces  a perturbation to the phase of the high-$k$ first order mode at the last scattering surface which is proportional to the wavenumber itself, and that therefore grows for more and more squeezed triangles. This growth however does not persist until very high $k$, because there is an averaging over the width of the last scattering surface, and we are left with an $f_{\rm NL}^{\rm eff.}$ of order a couple. 

A second effect due to the time delay is that it changes the amount of time during which photons can diffuse. This second effect goes together with another effect, the perturbation to the diffusion damping scale $k_D^2\propto n_e$ due to a low-$k$ $\delta_e$ mode. In the limit in which the $\delta_e$ perturbation is much slower that the timescale of recombination (which is approximately the case for low-$k$ modes), the enhanced effect from  the perturbation to $n_e$ is cancelled by the change in the  time during which the photons diffuse, as expected.
However, what is left of the perturbations is boosted in the squeezed limit by the square of ratio of the $k$ of the high-$k$ mode to the damping scale $k_D$, and again we are left with an effective $f_{\rm NL}^{\rm eff}$ of a few.

By summing all of these effects, we get our final signal corresponding to $f_{\rm NL}^{\rm eff.}\simeq-3.5\pm{\cal{O}}(1)$. Notice that this sign of $f_{\rm NL}^{\rm eff}$ corresponds to enhancing the short scale power in hotter long scale regions. Though physically well defined, experimentally quite large, and possibly detectable, the effect we find is nevertheless not much larger than the naive expectations for the effects coming from the terms we have neglected, which are expected to give rise to $f_{\rm NL}^{\rm eff.}$ of order one. We can not therefore exclude that other sources might give comparable signal. In particular it is not clear to us that other first order quantities that are not exponentially suppressed at high $k$ do not give rise to a similar effect. Therefore, in addition to providing a clear calculation of some of the second order effects, we think that our result further motivates the full calculation of the CMB anisotropies at second order.

\section*{Note added}
While this paper was being written, a similar paper \cite{Khatri:2008kb} appeared. Their calculation is quite similar to ours, though they do not include the second order monopole, dipole, and quadrupole in the collision term. For the bispectrum, this is equivalent to neglecting the perturbation $\delta_{k_D}$ to the diffusion damping due to $\delta n_e$, or in other words part $(a)$ of our source (see our sec.~\ref{sec: part a}). This plays a very important role  in our calculation and in the size of the total effect. Our analytical results from part $(b)$ of the source, though written in a different way,  agree with theirs (see our footnote \ref{footnote}).

\section*{Acknowledgments}

The work of LS is supported in part  by the National  Science Foundation under Grants No. PHY-0503584.

\begin{appendix}

\section{Rotations \label{appendix:A}}
We find it useful to summarize in this Appendix some of the transformation properties under the rotation group of the objects we treat in this paper. These formulas has been useful to do consistency
checks of our calculations.

Suppose we have a rotation $R$ that rotates the axis by
from $x,y,z$ to $x',y',z'$. A scalar transforms as:
\begin{equation}
\Delta'(\vec{k}',\hat n')=\Delta(R \vec k', R \hat n ')\ ,
\end{equation}
while a vector as
\begin{equation}
\vec v'(\vec k',\vec n')=R^{-1}\vec v( R \vec k',R\vec n')\ .
\end{equation}
The spherical harmonics transform according to the following relation
\begin{equation}
Y_{lm}(R\hat n')=D^l_{m,m'}(R^{-1})Y_{lm'}(\hat n')\ , 
\end{equation}
where $D^l(R)$ is the $2l+1\times 2l+1$ dimensional unitary matrix
representing the rotation $R$. It is also true that
\begin{equation}
D^l_{m0}(R_{\hat e_z\rightarrow \hat
k})=\sqrt{\frac{4\pi}{2l+1}}Y^*_{lm}(\hat k)\ ,
\end{equation}
\begin{equation}
D^{l}_{-m,-m'}(R){}^*=(-1)^{m-m'}D^l_{m,m'}(R)\ .
\end{equation}
Useful definitions of representations of the rotations $D^l$ and
some symmetry relations can be found in~\cite{Dsites}.

 Since
\begin{equation}
\Delta_{lm}(k)\sim \int d^2n Y^*_{lm}(\hat n) \Delta(\vec k,\hat
n)\ ,
\end{equation}
we have that in a rotated frame:
\begin{equation}\nonumber
\Delta'_{lm}(\vec k')\sim\int d^2n' Y^*_{lm}(\hat n')\Delta'(\hat
k',\hat n')\sim\int d^2n' Y^*_{lm}(\hat n')\Delta'(R\hat k,R\hat
n)\sim D^l_{m,m'}(R^{-1})\Delta_{lm'}(R\vec k') \ .
\end{equation}
So we have the property under rotations:
\begin{equation}
\Delta'_{lm}(\vec k')= D^l_{m,m'}(R^{-1})\Delta_{lm'}(R\vec k')\ .
\end{equation}
With these relationships we have verified that the equations in
sec. \ref{sec:second_order} and eq.~(\ref{eq: legendre_to_spherical}) transform correctly.

One can also perform a rotation of the integration variables in the
Gaunt integral and find the following nice relationship between 3-J symbols:
\begin{eqnarray}
\left(\begin{array}{ccc}
l_1 & l_2 & l_3 \\
m_1 & m_2 & m_3
\end{array}\right)
=D^{l_1}_{m_1',m_1}(R^{\pm 1}){}^{(*)}D^{l_2}_{m_2',m_2}(R^{\pm
1}){}^{(*)}D^{l_3}_{m_3',m_3}(R^{\pm
1}){}^{(*)}\left(\begin{array}{ccc}
l_1 & l_2 & l_3 \\
m_1' & m_2' & m_3'
\end{array}\right)\ ,
\end{eqnarray}
where ${}^{(*)}$ means that one can even take the complex
conjugate of that relation.

One further nice usage of these rotation properties is that they
allow us to compute scalar functions putting one of the $k$s on
the $z$ axis. We do not actually use this trick explicitly in the main part of the paper, but still, let us
show this works for the case of $\tilde
B_{l_1,l_2,l_3}(k_1,k_2,k_3)$, which is a scalar defined as
\begin{eqnarray}\label{eq:tildeb}
\tilde
B_{l_1,l_2,l_3}(k_1,k_2,k_3)=\sum_{m_1,m_2,m_3}\left(\begin{array}{ccc}
l_1 & l_2 & l_3 \\
m_1 & m_2 & m_3
\end{array}\right)\Delta_{l_1m_1}(k_1)\Delta_{l_2m_2}(k_2)\Delta_{l_3m_3}(k_3)\ .
\end{eqnarray}
(Notice that we do not even need to take the expectation value to be able to make use of the properties under rotations, this is the reason of the tilde symbol). We
have, applying a rotation (repeated indices are summed):
\begin{eqnarray}\label{eq: tilde B_1}
\tilde
B_{l_1,l_2,l_3}(k_1,k_2,k_3)&=&\left(\begin{array}{ccc}
l_1 & l_2 & l_3 \\
m_1 & m_2 & m_3
\end{array}\right) \\ \nonumber && D^{l_1}_{m_1,m_1'}(R^{-1})D^{l_2}_{m_2,m_2'}(R^{-1})D^{l_3}_{m_3,m_3'}(R^{-1})
\Delta_{l_1m_1'}'(R k_1)\Delta_{l_2m_2'}'(R
k_2)\Delta_{l_3m_3'}'(R k_3)\ .
\end{eqnarray}
Then we use that \cite{komatsu}:
\begin{equation}\nonumber
D^{l_1}_{m_1,m_1'}(R^{-1})D^{l_2}_{m_2,m_2'}(R^{-1})=\sum_{\tilde
l_3,\tilde m_3,\tilde m_3'} (2 \tilde l_3+1) D^{\tilde
l_3}_{\tilde m_3,\tilde m_3'}(R^{-1}){}^* \left(\begin{array}{ccc}
l_1 & l_2 & \tilde l_3 \\
m_1 & m_2 & \tilde m_3
\end{array}\right)
\left(\begin{array}{ccc}
l_1 & l_2 & \tilde l_3 \\
m_1' & m_2' & \tilde m_3'
\end{array}\right)\ ,
\end{equation}
and we substitute in (\ref{eq: tilde B_1}) and do the sum over
$m_1$ and $m_2$ using
\begin{eqnarray}
\sum_{m_1,m_2}\left(\begin{array}{ccc}
l_1 & l_2 & l_3 \\
m_1 & m_2 & m_3
\end{array}\right) \left(\begin{array}{ccc}
l_1 & l_2 & \tilde l_3 \\
m_1 & m_2 & \tilde m_3
\end{array}\right)=\frac{\delta_{l_3,\tilde l_3}\delta_{m_3,\tilde m_3}}{2
l_3+1}\ ,
\end{eqnarray}
obtaining:
\begin{eqnarray}
\tilde B_{l_1,l_2,l_3}(k_1,k_2,k_3)=&&\sum_{m_3,\tilde
m_3'}D^{l_3}_{m_3,m_3'}(R^{-1})D^{ l_3}_{m_3,\tilde
m_3'}(R^{-1}){}^*\left(\begin{array}{ccc}
l_1 & l_2 & \tilde l_3 \\
m_1' & m_2' & \tilde m_3'
\end{array}\right)\\ \nonumber &&\Delta_{l_1m_1'}'(R k_1)\Delta_{l_2m_2'}'(R k_2)\Delta_{l_3m_3'}'(R
k_3)\ .
\end{eqnarray}
Since by unitarity:
\begin{equation}
D^{ l_3}_{m_3,\tilde m_3'}(R^{-1}){}^*=D^{ l_3}_{\tilde
m_3',m_3}(R)\ ,
\end{equation}
the sum over $m_3$ gives a $\delta_{m_3',\tilde m_3'}$, obtaining:
\begin{eqnarray}
\tilde B_{l_1,l_2,l_3}(k_1,k_2,k_3)=\sum_{m_1',m_2',
m_3'}\left(\begin{array}{ccc}
l_1 & l_2 & \tilde l_3 \\
m_1' & m_2' & \tilde m_3'
\end{array}\right)\Delta_{l_1m_1'}'(R k_1)\Delta_{l_2m_2'}'(R k_2)\Delta_{l_3m_3'}'(R
k_3) \ ,
\end{eqnarray}
i.e. the same expression in form as in (\ref{eq:tildeb}), but computed with the rotated
variables (as it should be for a scalar quantity). Imagine now to take the expectation value, which amounts to actually compute $B_{l_1,l_2,l_3}$. In general, there will be three 3-dimensional $\vec k$ integral. If
one chooses to put the second order perturbation on $\vec k_1$, and
then chooses the reference frame where $R \vec k_1= k_1 \hat e_z$,
then one can use the analytical results found in the frame where $\vec k$ is along
$\hat z$. Then, one integrates over $\hat{R k_2},\hat{R k_3}$, and
finally the integration on $\hat k_1$ is trivial, giving just $4\pi$.
With this technique, we have been able to reproduce the result of
eq.~(\ref{eq: B from _b}).

\end{appendix}

\end{document}